\documentclass[reprint,twocolumn,aps,prd,superscriptaddress,floatfix,nofootinbib]{revtex4-1}
\usepackage{hyperref}
\usepackage{amssymb}
\usepackage{amsmath}
\usepackage{graphicx}
\usepackage{multirow}
\usepackage{verbatim}
\usepackage{esint}

\newcommand{\beq}{\begin{equation}}
\newcommand{\eeq}{\end{equation}}

\newcommand{\ud}{\textrm{d}}

\newcommand{\xOne}{\texttt{x1}}
\newcommand{\xTwo}{\texttt{x2}}
\newcommand{\xThree}{\texttt{x3}}
\newcommand{\yOne}{\texttt{y1}}
\newcommand{\yTwo}{\texttt{y2}}
\newcommand{\yThree}{\texttt{y3}}

\begin{document}

\title{SENR/NRPy+: Numerical relativity in singular curvilinear coordinate systems}

\author{Ian Ruchlin}
\affiliation{Department of Mathematics, West Virginia University, Morgantown, West Virginia 26506, USA}
\author{Zachariah B. Etienne}
\affiliation{Department of Mathematics, West Virginia University, Morgantown, West Virginia 26506, USA}
\affiliation{Center for Gravitational Waves and Cosmology, West Virginia University, Chestnut Ridge Research Building, Morgantown, West Virginia 26505, USA}
\author{Thomas W. Baumgarte}
\affiliation{Department of Physics and Astronomy, Bowdoin College, Brunswick, Maine 04011, USA}

\date{\today}

\begin{abstract}
We report on a new open-source, user-friendly numerical relativity
code package called SENR/NRPy+. Our code extends previous
implementations of the BSSN reference-metric formulation to a much
broader class of curvilinear coordinate systems, making it ideally
suited to modeling physical configurations with approximate or exact
symmetries. In the context of modeling black hole dynamics, it is
orders of magnitude more efficient than other widely used open-source
numerical relativity codes. NRPy+ provides a Python-based interface in
which equations are written in natural tensorial form and output at
arbitrary finite difference order as highly efficient C code, putting
complex tensorial equations at the scientist's fingertips without the
need for an expensive software license. SENR provides the algorithmic
framework that combines the C codes generated by NRPy+ into a
functioning numerical relativity code. We validate against two other
established, state-of-the-art codes, and achieve excellent
agreement. For the first time---in the context of moving puncture
black hole evolutions---we demonstrate nearly exponential
convergence of constraint violation and gravitational waveform errors
to zero as the order of spatial finite difference derivatives is
increased, while fixing the numerical grids at moderate resolution in
a singular coordinate system. Such behavior outside the horizons is
remarkable, as numerical errors do not converge to zero near punctures, and all
points along the polar axis are coordinate singularities. The
formulation addresses such coordinate singularities via cell-centered
grids and a simple change of basis that analytically regularizes
tensor components with respect to the coordinates. Future plans
include extending this formulation to allow dynamical coordinate grids
and bispherical-like distribution of points to efficiently capture
orbiting compact binary dynamics.
\end{abstract}


\maketitle

\section{Introduction}
\label{sec:introduction}

The Laser Interferometer Gravitational-Wave Observatory (LIGO) and Virgo
Scientific Collaboration's direct detections of six gravitational wave
events---five black hole binary mergers~\cite{Abbott:2016a,
  Abbott:2016b, Abbott:2017a, Abbott:2017b, Abbott:2017e} and one
neutron star binary merger~\cite{Abbott:2017c, Abbott:2017d}---have
ushered in the age of gravitational wave astrophysics and
multimessenger astronomy. As the signal-to-noise ratios of
gravitational wave detections grow with increased interferometer
sensitivity, the need to continually improve our theoretical models
of these phenomena is critical, as new physics may otherwise be
missed.

Just over a decade ago, breakthroughs in numerical 
relativity~\cite{Pretorius:2005a, Campanelli:2006, Baker:2006} 
opened the door to simulating the inspiral, merger, and ringdown 
phases of black hole binaries in vacuum. Black hole simulations---and, indeed,
all compact binary simulations---span many orders of magnitude
both in length scale and timescale. Making
them computationally tractable for reliable gravitational wave
predictions requires that the underlying numerical grid structure be
tuned to optimally sample the space. Ordered meshes that map to
Cartesian grids are most practical, as this greatly simplifies
algorithms for high-order approximations of spatial derivatives in
Einstein's equations.

There are currently two basic approaches in numerical relativity to
setting up numerical grids for compact binary simulations. The most
popular is to apply adaptive mesh refinement (AMR), where the grids consist
of nested Cartesian coordinate boxes at different, discrete numerical
resolutions. This enables the application of highest numerical
resolution where it is most needed: the strongly curved spacetime fields
inside and around compact objects. The most widely used AMR
infrastructure is provided by the open-source Cactus/Carpet
code~\cite{Goodale:2002, Schnetter:2004, Cactus, Carpet} 
within the Einstein Toolkit~\cite{Loffler:2012, EinsteinToolkit}
(ETK). One downside to this approach is that the compact objects of
interest are typically round, not rectangular, which results in a highly
inefficient distribution of grid points inside and near compact
objects (see, e.g.,~\cite{SENR:web} for a more detailed analysis). In
addition, the sudden change in grid resolution at the refinement
boundaries can produce spurious reflections in sharp gauge modes or
high-frequency gravitational wave features when using the moving puncture
formalism, resulting in poor convergence in gravitational waveforms
extracted from these numerical models~\cite{Zlochower:2012, Etienne:2014}.

The current alternative to adaptive mesh refinement in numerical
relativity codes, pioneered by the Simulating eXtreme Spacetimes (SXS)
Collaboration~\cite{SPEC} in their Spectral Einstein Code
(SpEC~\cite{Szilagyi:2014}, and its in-development successor
SpECTRE~\cite{Kidder:2017}), is to smoothly juxtapose a large number
of curvilinear three-dimensional grid patches, each with a smooth one-to-one
mapping to a Cartesian grid. Evaluating derivatives with these grids
requires computation of Jacobians, which, alongside managing the
dynamics of the grid structure itself, remains a significant
contributor to their codes' computational costs for compact binary
inspirals that are aimed at generating gravitational wave
predictions. 

Both approaches involve control systems that adjust the grids to track
compact objects as they orbit, although the SXS control system is far
more complex because their formalism~\cite{Pretorius:2005b}
additionally requires for stability that black hole interiors be very
carefully excised from the computational domain. Negative side effects
of the algorithmic complexity for both methods include a steep learning curve
for new users and difficulty in interpreting numerical errors.

We address these drawbacks by developing
a new code that builds on an innovative rescaling approach for solving
Einstein's equations in spherical coordinates. The approach is
designed to take full advantage of symmetries in the underlying
configuration, requiring as few numerical grid points as possible and
unlocking the desktop as a powerful tool for numerical relativity. Our
work builds on strategies that, in the context of spherical coordinates,
rescale tensors component by component so that the detrimental effects
of coordinate singularities on numerics are completely
removed~\cite{Bonazzola:2004, Shibata:2004, Brown:2009, Gourgoulhon:2012, Montero:2012, Baumgarte:2013}.
Treating all
coordinate singularities analytically, the equations can then be
integrated numerically without encountering instabilities. We
generalize this approach to a much broader class of static orthogonal
coordinate systems by absorbing the coordinate singularities out of
the tensor components and into a noncoordinate basis. The rescaling
strategy is implemented in the context of the BSSN reference-metric
formulation to enable highly efficient puncture black hole evolutions
using the moving puncture 
approach~\cite{Alcubierre:2003b, Campanelli:2006, Baker:2006} in a broad 
class of spherical-, cylindrical-, and Cartesian-like coordinate systems 
without special integration methods or introducing gaps in the numerical grid.

We implement this approach within a new, open-source code package
called SENR/NRPy+~\cite{SENR:web}. At its core, SENR/NRPy+ aims to be
as algorithmically simple and user friendly as 
possible, all while being highly efficient. In short,
SENR/NRPy+ aims to minimize both human and computational expense while
maximizing science outcomes.

SENR/NRPy+ is built upon the philosophy that the distribution of
points on the numerical grid should take maximum advantage of
approximate symmetries in the physical system. Compact object systems of
interest in gravitational wave astronomy typically possess a high
degree of angular symmetry, making spherical- and, more
generally, cylindrical-like coordinate systems ideal candidates for
efficient sampling.

NRPy+ (``Python-based code generation for numerical relativity and beyond'')
is designed to convert the BSSN reference-metric formulation
of the Einstein equations, in a broad class of orthogonal coordinate
systems, from Einstein-like notation directly into C code. It operates
without the need for expensive, proprietary computer algebra systems
like Mathematica or Maple. As its name suggests, NRPy+ is based
entirely in Python and depends only on the standard Python computer
algebra package SymPy~\cite{Meurer:2017} for symbolic algebra, which
is widely available on supercomputing clusters.

SENR (``the Simple, Efficient Numerical Relativity code'')
incorporates C codes generated by NRPy+ to form a complete,
OpenMP-parallelized~\cite{Dagum:1998} numerical relativity code. Its
skeletal structure makes the algorithmic underpinnings of numerical
relativity codes transparent to the user. 

We verify SENR/NRPy+ by direct comparisons with two other numerical
relativity codes that are both well established in the literature. In
the context of strongly perturbed Minkowski spacetime 
(a version of the robust stability test~\cite{Szilagyi:2000,
  Szilagyi:2002,Alcubierre:2004,ApplesRobustStabilityTest}), we
achieve roundoff-level agreement with the Baumgarte \emph{et al.}~\cite{Baumgarte:2013} 
code, which evolves the BSSN equations in
spherical coordinates at fixed fourth-order finite difference
accuracy. We also demonstrate excellent agreement between the results
of SENR/NRPy+ and the ETK in the context of simulating a
single, dynamical black hole. Then, we perform simulations of single-
and double-black-hole spacetimes, demonstrating that the
finite difference truncation error converges to zero with increasing
grid resolution at the expected rate.

Perhaps most importantly, we show for the first time that---in the
context of moving puncture evolutions---the
truncation error in our finite differencing scheme converges to
zero nearly exponentially\footnote{When finite difference truncation error dominates,
  we expect numerical error to scale approximately as
  $\left|C_n u^{(n+1)}(\xi)\right|(\Delta x)^n$, where $n = N_{\text{FD}}$ is the finite
  difference order, $\xi$ is in the neighborhood of the point at which
  we evaluate the derivative of the function $u$, and $C_n \sim 1/4^n$
  for a centered stencil on a numerical
  grid with uniform spacing $\Delta x$. In the case that $\left|u^{(n+1)}(\xi)\right|$ is bounded
  (e.g.,~$u(r)=\sin r$), pure exponential
  convergence of the finite difference derivative error is observed as
  $n$ is increased and $\Delta x$ is held fixed (again, assuming that
  truncation error dominates). When simulating gravitational fields,
  $\left|u^{(n+1)}(\xi)\right|$ can grow as $n!$ (e.g.,~$u(r)=1/(1-r)$, $r \ne 1$),
  reducing the rate of exponential convergence at the
  finite difference orders we typically choose ($N_{\text{FD}} \in \{2,4,6,8,10\}$). We
  refer to this behavior as {\it nearly exponential convergence}.} outside puncture
black hole horizons with linear increase in the finite difference
order, keeping the numerical grids fixed at moderate
resolution.

The fact that we observe nearly exponential convergence is
remarkable for two reasons. First, the numerical grids chosen for
these simulations possess coordinate singularities at all points where
$r\sin(\theta)=0$. Therefore, near-exponential convergence in regions within
causal contact of these singularities demonstrates that our tensor
rescaling algorithm and cell-centered grids completely eliminate
convergence problems related to these coordinate
singularities. Second, a puncture black hole exhibits nonsmooth
fields at the site of the puncture, meaning we should have no {\it a
  priori} expectation of near-exponential numerical convergence
outside the black hole, either.
We attribute the observed convergence to the fact that the characteristics
of the physical (nongauge) fields, in the vicinity of the puncture, point
towards the puncture. With sufficient resolution inside of the 
horizon, the errors resulting from finite differencing across the puncture
singularity become trapped near the puncture, and are not able to escape
and contaminate the simulation at large~\cite{Brown:2008, Brown:2009b}.

Future SENR projects will involve further extending the formalism to
handle dynamical, bispherical-like coordinate systems, so that compact
binary dynamics may be modeled with minimum computational expense. In
this work we demonstrate near-exponential convergence of gravitational
waves with increased finite differencing order in the context of
head-on collisions of puncture black holes, and will build on this
success to tackle the orbital black hole binary problem as our next step.

This paper is organized as follows. In
Sec.~\ref{sec:evolution_equations}, we present the reference-metric
formulation of the BSSN and gauge
evolution equations. In Sec.~\ref{sec:tensor_rescaling}, we outline
the tensor component rescaling procedure that makes it possible to
evolve gravitational fields in singular coordinate systems. In
Sec.~\ref{sec:SENR}, we describe the structure of the SENR/NRPy+ code,
including the implementation of grid structures, coordinate system
options, diagnostics, and boundary conditions. In
Sec.~\ref{sec:results}, we first demonstrate that SENR/NRPy+ agrees with the
results of other established numerical relativity codes. Then, we show that
numerical errors converge to zero at the expected rates for a
nonspinning black hole with varying grid resolution or finite difference order,
and in different coordinate systems and gauges. Finally, in the
context of head-on collisions of two nonspinning puncture black holes,
we demonstrate near-exponential convergence of the gravitational
waveforms with increasing finite difference order. We conclude and
present plans for future work in Sec.~\ref{sec:Conclusion}.

Throughout this paper geometrized units are adopted, in which $c = 1$ and
$G = 1$. Latin indices $(i, j, k, \ldots)$ denote spatial degrees of
freedom and obey the Einstein summation convention.

\section{BSSN and Gauge Evolution Equations}
\label{sec:evolution_equations}

In this section, we describe our strategy for solving Einstein's equations, based on the
tensor-weight-zero BSSN~\cite{Nakamura:1987, Shibata:1995, Baumgarte:1999}
formulation of Brown~\cite{Brown:2009}.
Our numerical implementation extends the rescaling approach, developed by
Baumgarte \emph{et al.}~\cite{Baumgarte:2010} for spherical
coordinates, to a broader class of singular curvilinear coordinate
systems. In this paper, we focus on \mbox{spherical-}, \mbox{cylindrical-}, and
Cartesian-like coordinate systems, though the method can be easily
extended to many others.
For any of our coordinate grids, the reference metric $\hat{\gamma}_{i j}$
represents the flat space metric components expressed in a coordinate
basis.
In Sec.~\ref{sec:tensor_rescaling}, we use $\hat{\gamma}_{i j}$ in the rescaling procedure 
to define a noncoordinate basis,
in terms of which which all tensor components are explicitly free of coordinate singularities.
We assume that the background is independent of the coordinate time $t$, so that
$\partial_{t} \hat{\gamma}_{i j} = 0$. All hatted quantities are associated with the reference metric.

The reference metric is used to
decompose the conformal metric $\bar{\gamma}_{i j}$ into a correction
about the flat background 
\beq
  \label{eq:conformal_metric}
  \bar{\gamma}_{i j} = \hat{\gamma}_{i j} + \varepsilon_{i j} \; ,
\eeq
where the components in $\varepsilon_{i j}$ are not necessarily small
and contain the metric fields that are evolved on our numerical grids. 
The conformal metric is related to the physical 
spatial metric $\gamma_{i j}$ though a conformal rescaling
\beq
  \label{eq:physical_metric}
  \gamma_{i j} = e^{4 \phi} \bar{\gamma}_{i j} \; .
\eeq
Taking the determinant of~\eqref{eq:physical_metric} we observe that the 
conformal factor $e^{\phi}$ can be expressed as
\beq
  \label{eq:conf_factor}
  e^{\phi} = \left (\frac{\gamma}{\bar{\gamma}} \right )^{1/12} \; ,
\eeq
where $\gamma \equiv \det(\gamma_{i j})$ 
and $\bar{\gamma} \equiv \det(\bar{\gamma}_{i j})$ are the metric determinants. 
Quantities associated with the conformal metric $\bar{\gamma}_{i j}$ 
are barred. For example, the conformal covariant derivative operator
$\bar{D}_{i}$ is defined with respect to the conformal metric. The
inverse conformal metric $\bar{\gamma}^{i j}$ is defined to satisfy
\beq
  \bar{\gamma}^{i k} \bar{\gamma}_{k j} = \delta_{j}^{i} \; ,
\eeq
where $\delta_{j}^{i}$ is the Kronecker delta tensor.

The conformal rescaling~\eqref{eq:physical_metric} is not yet unique. In contrast to the 
original BSSN formulation, in which $\bar{\gamma}$ was set to unity, we adopt Brown's
``Lagrangian'' choice~\cite{Brown:2009}
\beq
  \label{eq:Lagrangian_detg}
  \partial_{t} \bar{\gamma} = 0 \; ,
\eeq
so that $\bar{\gamma}$ remains equal to its initial value. In particular, this implies that 
both $\gamma$ and $\bar{\gamma}$ are allowed to transform as determinants, 
i.e.,~as scalar densities of weight two. According to~\eqref{eq:conf_factor} the conformal factor 
$e^{\phi}$ then transforms as a scalar, rather than a scalar density, and
all other tensorial objects in our formalism similarly transform as tensors of weight zero (see 
also~\cite{Bonazzola:2004, Brown:2009}).
We choose ${\bar{\gamma} = \hat{\gamma} \equiv \det(\hat{\gamma}_{i j})}$
in the initial data for all applications in this
paper. Since both determinants remain independent of time, they remain
equal to each other throughout the evolution.

It is well known that Christoffel symbols do not transform covariantly
between coordinate systems. However, the difference of two sets of
Christoffel symbols is tensorial. We define the tensor
\beq
  \Delta_{j k}^{i} \equiv \bar{\Gamma}_{j k}^{i} - \hat{\Gamma}_{j k}^{i} \; ,
\eeq
whose indices are raised and lowered with the conformal metric.
It is useful to construct a vector by taking the trace
\beq
  \Delta^{i} \equiv \bar{\gamma}^{j k} \Delta_{j k}^{i} \; .
\eeq
In addition, the tensor-weight-zero conformal connection coefficient
three-vector $\bar{\Lambda}^{i}$ is evolved independently, and satisfies the initial
constraint
\beq
  \label{eq:conformal_connection_constraint}
  \mathcal{C}^{i} \equiv \bar{\Lambda}^{i} - \Delta^{i} = 0 \; .
\eeq

The conformal, trace-free part of the extrinsic curvature is denoted
\beq
  \bar{A}_{i j} = e^{-4 \phi} \left (K_{i j} - \frac{1}{3} \gamma_{i j} K \right ) \; ,
\eeq
where $K_{i j}$ is the physical extrinsic curvature and 
${K = \gamma^{i j} K_{i j}}$ is the mean curvature.

Defining the hypersurface-normal derivative operator
\beq
  \partial_{\perp} \equiv \partial_{t} - \mathcal{L}_{\beta} \; ,
\eeq
where $\mathcal{L}_{\beta}$ is the Lie derivative along the shift
vector $\beta^{i}$, the BSSN evolution system in vacuum is written
as~\cite{Brown:2009}
\begin{subequations}
  \label{eq:BSSN}
\begin{align}
  \label{eq:metric_RHS}
  \partial_{\perp} \varepsilon_{i j} {} = {} & \frac{2}{3} \bar{\gamma}_{i j} \left (\alpha \bar{A}_{k}^{k} - \bar{D}_{k} \beta^{k}\right ) + 2 \hat{D}_{(i} \beta_{j)} - 2 \alpha \bar{A}_{i j} \; , \\
  \partial_{\perp} \bar{A}_{i j} {} = {} & -\frac{2}{3} \bar{A}_{i j} \bar{D}_{k} \beta^{k} - 2 \alpha \bar{A}_{i k} {\bar{A}^{k}}_{j} + \alpha \bar{A}_{i j} K \nonumber \\
  & + e^{-4 \phi} \left \{-2 \alpha \bar{D}_{i} \bar{D}_{j} \phi + 4 \alpha \bar{D}_{i} \phi \bar{D}_{j} \phi \right . \nonumber \\
    & \left . + 4 \bar{D}_{(i} \alpha \bar{D}_{j)} \phi - \bar{D}_{i} \bar{D}_{j} \alpha + \alpha \bar{R}_{i j} \right \}^{\text{TF}} \; , \\
  \partial_{\perp} W {} = {} & -\frac{1}{3} W \left (\bar{D}_{k} \beta^{k} - \alpha K \right ) \; , \\
  \partial_{\perp} K {} = {} & \frac{1}{3} \alpha K^{2} + \alpha \bar{A}_{i j} \bar{A}^{i j} \nonumber \\
  & - e^{-4 \phi} \left (\bar{D}_{i} \bar{D}^{i} \alpha + 2 \bar{D}^{i} \alpha \bar{D}_{i} \phi \right ) \; , \\
  \partial_{\perp} \bar{\Lambda}^{i} {} = {} & \bar{\gamma}^{j k} \hat{D}_{j} \hat{D}_{k} \beta^{i} + \frac{2}{3} \Delta^{i} \bar{D}_{j} \beta^{j} + \frac{1}{3} \bar{D}^{i} \bar{D}_{j} \beta^{j} \nonumber \\
  & - 2 \bar{A}^{i j} \left (\partial_{j} \alpha - 6 \partial_{j} \phi \right ) + 2 \bar{A}^{j k} \Delta_{j k}^{i} \nonumber \\
  & -\frac{4}{3} \alpha \bar{\gamma}^{i j} \partial_{j} K \; .
\end{align}
\end{subequations}
In the above, ``TF'' denotes the trace-free part of the expression in
brackets, the conformal factor is evolved as $W = e^{-2 \phi}$
(following, e.g.,~\cite{Marronetti:2008, Montero:2012} to ensure
smoother spacetime fields near puncture black holes), and the
components of the conformal Ricci tensor are calculated by
\begin{align}
  \bar{R}_{i j} {} = {} & - \frac{1}{2} \bar{\gamma}^{k l} \hat{D}_{k} \hat{D}_{l} \bar{\gamma}_{i j} + \bar{\gamma}_{k(i} \hat{D}_{j)} \bar{\Lambda}^{k} + \Delta^{k} \Delta_{(i j) k} \nonumber \\
  & + \bar{\gamma}^{k l} \left (2 \Delta_{k(i}^{m} \Delta_{j) m l} + \Delta_{i k}^{m} \Delta_{m j l} \right ) \; .
\end{align}
The trace-free condition $\bar{\gamma}^{i j} \bar{A}_{i j} = 0$, which
can be violated by numerical error, is enforced dynamically by the
term proportional to $\bar{A}_{i}^{i}$ in
Eq.~\eqref{eq:metric_RHS}~\cite{Brown:2009}. In this paper, we
restrict ourselves to vacuum spacetimes so that all of the matter
source terms vanish.

The evolution system is completed with specification of the gauge: the
lapse function $\alpha$ and the shift vector $\beta^{i}$. Unless
otherwise stated, we employ the advective 1+log lapse
condition~\cite{Bona:1995}
\beq
  \label{eq:1_plus_log}
  \partial_{0} \alpha = -2 \alpha K
\eeq
and the advective Gamma-driver shift condition~\cite{Alcubierre:2003b}
\begin{subequations}
  \label{eq:Gamma-driver}
\begin{align}
  \partial_{0} \beta^{i} &= B^{i} \; , \\
  \label{eq:Gamma-driver_B}
  \partial_{0} B^{i} &= \frac{3}{4} \partial_{0} \bar{\Lambda}^{i} - \eta B^{i} \; .
\end{align}
\end{subequations}
Here, $B^{i}$ is an auxiliary vector, $\eta$ is a (dimensionful)
damping parameter~\cite{Schnetter:2010}, and the
(noncovariant) advective derivative operator is defined as 
(see Case No.\ 8 in Table I of~\cite{vanMeter:2006})
\beq
  \partial_{0} \equiv \partial_{t} - \beta^{j} \partial_{j} \; .
  \label{eq:partial0}
\eeq
The BSSN equations coupled to
these gauge conditions are known together as the moving puncture
approach~\cite{Alcubierre:2003b, Campanelli:2006, Baker:2006}.
A total of 24 fields are evolved.

\section{Tensor Rescaling}
\label{sec:tensor_rescaling}

In the previous section, we described the reference-metric formulation
of the BSSN evolution equations in the moving puncture
approach~\eqref{eq:BSSN},~\eqref{eq:1_plus_log},
and~\eqref{eq:Gamma-driver}. These equations
form the foundation for solving Einstein's equations in any coordinate
system we like, and we leverage this coordinate freedom to take
maximum advantage of near symmetries in physical systems. While, from
an analytic perspective, there is no problem solving Einstein's
equations in arbitrary coordinate systems, numerical solutions diverge
if the chosen coordinate system possesses a coordinate singularity.

Well-known examples of coordinate singularities include the points at
$\rho = 0$ in cylindrical coordinates or at $r \sin(\theta) = 0$ in
spherical coordinates. Tensor components that are regular everywhere in a
Cartesian basis will inherit these singularities during the change of
basis to those coordinates. The fact that these singularities are a
consequence of the coordinates themselves, and not of the underlying
tensor fields, means that they can in principle be scaled out of the
tensor components in a way that is consistent with the adopted
reference-metric formulation of BSSN.

The goal of the tensor rescaling is to analytically absorb singular terms
into the noncoordinate basis. Tensor components are
naturally regular with respect to the coordinates when expressed in
terms of the noncoordinate basis. 

In Sec.~\ref{sec:spherical_rescaling}, we show a simple example of rescaling a rank-1
tensor with one vector component and a rank-2 tensor with one
two-dual-vector component in the case of ordinary spherical
coordinates. We generalize the rescaling procedure to arbitrary
coordinate distributions in Sec.~\ref{sec:general_rescaling}.

\subsection{Spherical rescaling examples}
\label{sec:spherical_rescaling}

In this section, we refer to objects in Cartesian coordinates with
indices $i$ and $j$, and to objects in spherical coordinates with
indices $a$ and $b$.

Consider a finite vector field $\boldsymbol{V}$ with
components $V^{i}$ in Cartesian coordinates $y^{(i)} = (x, y, z)$ and a Cartesian coordinate
basis $\frac{\boldsymbol{\partial}}{\boldsymbol{\partial} y^{(i)}}$
\beq
  \boldsymbol{V} = V^{i} \frac{\boldsymbol{\partial}}{\boldsymbol{\partial} y^{(i)}} \equiv V^{x} \frac{\boldsymbol{\partial}}{\boldsymbol{\partial} x} + V^{y} \frac{\boldsymbol{\partial}}{\boldsymbol{\partial} y} + V^{z} \frac{\boldsymbol{\partial}}{\boldsymbol{\partial} z} \; .
\eeq
An index in parentheses labels the individual coordinate functions or basis vectors, not
vector components. For simplicity, and without loss of generality, we
consider a vector that possesses only one nontrivial component 
($V^{y} = V^{z} = 0$)
\beq
  \boldsymbol{V} = V^{x} \frac{\boldsymbol{\partial}}{\boldsymbol{\partial} x} \; .
\eeq
The component $V^{x}$ is a smooth, finite function of the
Cartesian coordinate values.

In Cartesian coordinates, the natural set of noncoordinate basis
vectors $\boldsymbol{e}_{(i)}$ coincides with the coordinate basis,
and therefore
\beq
  \boldsymbol{V} = V^{x} \boldsymbol{e}_{(x)} \; .
\eeq
Now, we transform the vector to
ordinary, uniform spherical coordinates $r^{(a)} = (r, \theta, \varphi)$,
related to the Cartesian coordinates by
\begin{subequations}
  \label{eq:Cartesian_spherical}
\begin{align}
  x &= r \sin(\theta) \cos(\varphi) \; , \\
  y &= r \sin(\theta) \sin(\varphi) \; , \\
  z &= r \cos(\theta) \; .
\end{align}
\end{subequations}
This coordinate relationship characterizes the familiar Jacobian
matrix with components
\beq
  \label{eq:spherical_jacobian}
  \frac{\partial y^{i}}{\partial r^{a}} = \begin{pmatrix}
    \sin(\theta) \cos(\varphi) & r \cos(\theta) \cos(\varphi) & -r \sin(\theta) \sin(\varphi) \\
    \sin(\theta) \sin(\varphi) & r \cos(\theta) \sin(\varphi) & r \sin(\theta) \cos(\varphi) \\
    \cos(\theta) & -r \sin(\theta) & 0
  \end{pmatrix}
\eeq
and the inverse Jacobian matrix 
$\frac{\partial r^{a}}{\partial y^{i}}$ satisfies
\beq
  \frac{\partial y^{i}}{\partial r^{a}} \frac{\partial r^{a}}{\partial y^{j}} = \delta_{j}^{i} \; , \quad \frac{\partial y^{i}}{\partial r^{b}} \frac{\partial r^{a}}{\partial y^{i}} = \delta_{b}^{a} \; .
\eeq
An application of the ordinary derivative chain rule yields the
transformation formula for the coordinate basis vectors
\beq
  \label{eq:chain_rule}
  \frac{\boldsymbol{\partial}}{\boldsymbol{\partial} y^{(i)}} = \frac{\partial r^{a}}{\partial y^{i}} \frac{\boldsymbol{\partial}}{\boldsymbol{\partial} r^{(a)}} \; .
\eeq
The Cartesian components $V^{i}$ are transformed from the spherical
coordinate components $V^{a}$ using the inverse Jacobian
\beq
  V^{i} = V^{a} \frac{\partial y^{i}}{\partial r^{a}} \; .
\eeq
It is this fact, that a tensor's components transform in a way that is
inverse to the basis, which preserves the geometric meaning of a
tensor field represented in any coordinate system. The resulting
vector components in the spherical coordinate basis are
\begin{align}
  \label{eq:example_spherical_vector}
  \boldsymbol{V} = {} & V^{a} \frac{\boldsymbol{\partial}}{\boldsymbol{\partial} r^{(a)}} = V^{x} \frac{\partial r^{a}}{\partial x} \frac{\boldsymbol{\partial}}{\boldsymbol{\partial} r^{(a)}} \nonumber \\
  = {} & V^{x} \sin(\theta) \cos(\varphi) \frac{\boldsymbol{\partial}}{\boldsymbol{\partial} r} + \frac{V^{x} \cos(\theta) \cos(\varphi)}{r} \frac{\boldsymbol{\partial}}{\boldsymbol{\partial} \theta} \nonumber \\
  &- \frac{V^{x} \sin(\varphi)}{r \sin(\theta)} \frac{\boldsymbol{\partial}}{\boldsymbol{\partial} \varphi} \; .
\end{align}
The coordinate singularities that have been introduced by the Jacobian
become obvious when the components are evaluated at the origin $r = 0$
and along the polar axis $\sin(\theta) = 0$. 

We absorb this undesirable behavior into the basis vectors as
follows. For the noncoordinate spherical basis $\boldsymbol{e}_{(a)}$,
choose orthogonal vectors
\begin{subequations}
\label{eq:spherical_vector_basis}
\begin{align}
  \boldsymbol{e}_{(r)} &= \frac{\boldsymbol{\partial}}{\boldsymbol{\partial} r} \; , \\
  \boldsymbol{e}_{(\theta)} &= \frac{1}{r} \frac{\boldsymbol{\partial}}{\boldsymbol{\partial} \theta} \; , \\
  \boldsymbol{e}_{(\varphi)} &= \frac{1}{r \sin(\theta)} \frac{\boldsymbol{\partial}}{\boldsymbol{\partial} \varphi} \; .
\end{align}
\end{subequations}
In terms of these, the vector components become
\begin{align}
  \label{eq:example_spherical_vector_rescaled}
  \boldsymbol{V} = {} & V^{x} \sin(\theta) \cos(\varphi) \boldsymbol{e}_{(r)} + V^{x} \cos(\theta) \cos(\varphi) \boldsymbol{e}_{(\theta)} \nonumber \\
  &- V^{x} \sin(\varphi) \boldsymbol{e}_{(\varphi)} \; ,
\end{align}
which are manifestly regular over the entire domain.

To demonstrate that this strategy is more generally applicable, consider a
rank-2 tensor $\boldsymbol{W}$ with only one nontrivial component in a
Cartesian coordinate two-dual-vector basis
\beq
  \boldsymbol{W} = W_{x x} \, \mathbf{d} x \otimes \mathbf{d} x \; ,
\eeq
where $\otimes$ is the tensor product and $\mathbf{d}$ is the exterior
derivative operator acting on the scalar coordinate functions to
produce one-forms $\mathbf{d} y^{(i)}$ (i.e.\ dual vectors). The
transformation rule dual to~\eqref{eq:chain_rule} is
\beq
  \mathbf{d} y^{(i)} = \frac{\partial y^{i}}{\partial r^{a}} \, \mathbf{d} r^{(a)} \; .
\eeq

Two contractions with the inverse Jacobian matrix transforms the
tensor components to the spherical coordinate basis
\begin{align}
  \boldsymbol{W} = {} & W_{x x} \sin^{2}(\theta) \cos^{2}(\varphi) \, \mathbf{d} r \otimes \mathbf{d} r \nonumber \\
  &+ 2 W_{x x} r \sin(\theta) \cos(\theta) \cos^{2}(\varphi) \, \mathbf{d} r \otimes \mathbf{d} \theta \nonumber \\
  &- 2 W_{x x} r \sin^{2}(\theta) \sin(\varphi) \cos(\varphi) \, \mathbf{d} r \otimes \mathbf{d} \varphi  \nonumber \\
  &+ W_{x x} r^{2} \cos^{2}(\theta) \cos^{2}(\varphi) \, \mathbf{d} \theta \otimes \mathbf{d} \theta \nonumber \\
  &- 2 W_{x x} r^{2} \sin(\theta) \cos(\theta) \sin(\varphi) \cos(\varphi) \, \mathbf{d} \theta \otimes \mathbf{d} \varphi \nonumber \\
  &+ W_{x x} r^{2} \sin^{2}(\theta) \sin^{2}(\varphi) \, \mathbf{d} \varphi \otimes \mathbf{d} \varphi \; .
\end{align}
Notice that the components of $\boldsymbol{W}$ vanish when evaluated
at, say, $r = 0$ and $\theta = \pi$, which amounts to a coordinate
singularity that destroys information about the tensor's value in
other bases (e.g., Cartesian) at these points. This singular behavior
is dual to that seen above, where the component values of
$\boldsymbol{V}$ in the spherical coordinate basis became unbounded
at certain locations. 

Again, this is ameliorated by an alternative choice of basis. The
noncoordinate spherical basis one-forms are defined as the dual to the
basis vectors~\eqref{eq:spherical_vector_basis} 
\begin{subequations}
\begin{align}
  \boldsymbol{e}^{(r)} &= \mathbf{d} r \; , \\
  \boldsymbol{e}^{(\theta)} &= r \, \mathbf{d} \theta \; , \\
  \boldsymbol{e}^{(\varphi)} &= r \sin(\theta) \, \mathbf{d} \varphi \; .
\end{align}
\end{subequations}
In this noncoordinate basis, the tensor components become
\begin{align}
  \boldsymbol{W} = {} & W_{x x} \sin^{2}(\theta) \cos^{2}(\varphi) \boldsymbol{e}^{(r)} \otimes \boldsymbol{e}^{(r)}  \nonumber \\
  &+ 2 W_{x x} \sin(\theta) \cos(\theta) \cos^{2}(\varphi) \boldsymbol{e}^{(r)} \otimes \boldsymbol{e}^{(\theta)} \nonumber \\
  &- 2 W_{x x} \sin(\theta) \sin(\varphi) \cos(\varphi) \boldsymbol{e}^{(r)} \otimes \boldsymbol{e}^{(\varphi)} \nonumber \\
  &+ W_{x x} \cos^{2}(\theta) \cos^{2}(\varphi) \boldsymbol{e}^{(\theta)} \otimes \boldsymbol{e}^{(\theta)} \nonumber \\
  &- 2 W_{x x} \cos(\theta) \sin(\varphi) \cos(\varphi) \boldsymbol{e}^{(\theta)} \otimes \boldsymbol{e}^{(\varphi)} \nonumber \\
  &+ W_{x x} \sin^{2}(\varphi) \boldsymbol{e}^{(\varphi)} \otimes \boldsymbol{e}^{(\varphi)} \; .
\end{align}
Now, for any point in the spherical domain, it can be shown that all
of the above components vanish simultaneously if and only if $W_{x x} = 0$. 

For both $\boldsymbol{V}$ and $\boldsymbol{W}$---as well as higher
rank tensors in general---these arguments extend to arbitrary
Cartesian tensors, allowing all components to be nontrivial.

\subsection{The general rescaling procedure}
\label{sec:general_rescaling}

The rescaling examples reviewed in the previous section can be
generalized by treating the noncoordinate basis as a collection of
matrix operators (and the dual basis as the associated inverse
operators), and using them to project the singularities out of tensor
components represented in a coordinate basis. Since the difference
between bases is only a coordinate transformation, and the BSSN
formulation we adopt is covariant, we are free to apply this strategy
to remove coordinate singularities from tensorial expressions in the
formulation. This section reviews the general procedure.

We denote the noncoordinate basis vectors by $e_{(j)}^{i}$, where the index $(j)$
lists the individual basis vectors and $i$ labels the components of a
particular vector with respect to the underlying coordinate basis.
By definition, the set of basis vectors $\{\boldsymbol{e}_{(i)}\}$ is linearly
independent and spans the tangent space at every point in the spatial
hypersurface. We restrict our consideration to time-independent,
orthonormal bases. There exists a dual basis $e_{j}^{(i)}$ satisfying
\beq
  \label{eq:dual_basis}
  e_{k}^{(i)} e_{(j)}^{k} = e_{j}^{(k)} e_{(k)}^{i} = \delta_{j}^{i} \; .
\eeq
The components of the flat background reference metric, represented in a
coordinate basis, are related algebraically to the basis dual vectors via
\beq
  \label{eq:basis_definition}
  \hat{\gamma}_{i j} = \delta_{k l} e^{(k)}_{i} e^{(l)}_{j} \; .
\eeq
In this way, the reference metric is constructed as the ``product'' of basis 
vectors, or, equivalently, that the basis constitutes the ``square root'' of the reference
metric~\cite{Carroll:2004}. We treat this relationship as the
definition of the noncoordinate basis components in terms of the known
flat space reference-metric components in the corresponding coordinate
basis. The noncoordinate basis, defined as such, is an orthonormal basis.
The components $e_{j}^{(i)}$ are sometimes referred to as the ``scale
factors'' of the reference metric, and they contain the singularities associated
with the coordinates. The components of other tensors are made regular 
with respect to the coordinates by factoring out the scale factors in 
the appropriate way. For example~\eqref{eq:conformal_metric}
\beq
  \varepsilon_{i j} = h_{k l} e^{(k)}_{i} e^{(l)}_{j}
\eeq
and~\eqref{eq:conformal_connection_constraint}
\beq
  \bar{\Lambda}^{i} = \bar{\lambda}^{j} e_{(j)}^{i} \; .
\eeq
The rescaled components are recovered by the inverse relationships
\beq
  h_{i j} = \varepsilon_{k l} e^{k}_{(i)} e^{l}_{(j)}
\eeq
and
\beq
  \bar{\lambda}^{i} = \bar{\Lambda}^{j} e_{j}^{(i)} \; .
\eeq
The rescaled tensor components $h_{i j}$, $\bar{\lambda}^{i}$, and so
on, are regular with respect to the coordinate singularities. (Note
that $h_{i j}$ and $\bar{\lambda}^{i}$ in spherical coordinates
correspond directly to the functions of the same name
in~\cite{Baumgarte:2013}.) The BSSN evolution equations~\eqref{eq:BSSN}
are tensorial, and are therefore independent of the choice of basis.

The rescaled fields are those that
are integrated and differentiated numerically on the coordinate grid,
described next, in Sec.~\ref{sec:SENR}. This rescaling procedure enables us to
achieve stable and convergent solutions in a broad class of singular coordinate
systems, as we demonstrate in Sec.~\ref{sec:results}.

\section{The SENR/NRPy+ Code}
\label{sec:SENR}

Numerical relativity codes built to evolve 3+1 initial value
formulations of Einstein's equations generally contain thousands
of lines of code just to express the needed equations for initial
data, time evolution, and diagnostics. Early incarnations were largely
coded by hand, exacerbating the already laborious and time-consuming
debugging process. Most numerical relativity groups have migrated to
automatic code generation, typically relying on closed-source,
proprietary computer algebra systems like Maple or Mathematica to
directly convert tensorial expressions typed by hand directly in
  Einstein-like notation into, e.g., highly optimized C
code. Kranc~\cite{Husa:2006} is one example of a very nice
open-source, Mathematica-based package for converting
Einstein's equations---written in Einstein-like notation---into optimized
C code.

Proprietary packages like Mathematica or Maple require
expensive licenses that some users simply cannot afford, creating a
barrier to entry for potential developers. Further, most numerical
relativity simulations are performed with supercomputing systems, on
which licenses for, e.g., Mathematica or Maple are not available---again
due to the high licensing cost. As a workaround, numerical relativists
will often generate their code locally, using Mathematica or Maple, and then
transfer it to the supercomputing system---just
another way that these licenses can inconvenience users.

NRPy+ (``Python-based code generation for numerical relativity and
beyond'') aims to address these issues. It is the first
open-source~\cite{BSDLicense, GPLLicense},
non-Mathematica- or Maple-based code generation package for tensorial
expressions written in Einstein notation. NRPy+ is written entirely in
Python\footnote{Both Python 2.7+ and 3.0+ are supported.} and depends
only on the standard Python computer algebra package
SymPy~\cite{Meurer:2017} for symbolic algebra, which is widely
available on supercomputing clusters.

If we wish to solve Einstein's equations in a new coordinate system
with NRPy+, we need only define the corresponding reference metric in
terms of its scale factors. Using these as input, NRPy+
generates Einstein's equations in these coordinates and outputs
OpenMP-capable C code that is highly optimizable (SIMD vectorized) by
compilers, resulting in a tremendous performance boost compared to
simple serial implementations. NRPy+ also leverages SymPy's
ability to eliminate common subexpressions from complicated algebraic
expressions, minimizing the number of floating point operations per
expression evaluation. 

SENR (``the Simple, Efficient Numerical Relativity code'') is a
complete, OpenMP-parallelized~\cite{Dagum:1998} numerical relativity
code, incorporating the C codes generated by NRPy+ wherever complicated
tensorial expressions are needed. Its skeletal structure makes the
algorithms on which numerical relativity codes are based transparent
to the user.

The division of labor between SENR and NRPy+ is outlined in
Table~\ref{tab:SENR_NRPy}, providing a convenient launching
point for later subsections that expand on this structure.

\begin{table*}[!t]
  \centering
  \caption{The division of labor between the SENR C code and the NRPy+
    Python code, with a link to the relevant section detailing each task.
    To perform a simulation, NRPy+ is first run to
    automatically generate C files containing necessary initial data,
    evolution, and diagnostic equations, coupled to highly optimized
    finite difference codes as needed for spatial derivatives. SENR
    contains all of the infrastructure needed to make use of these C
    codes in the context of a full numerical simulation, complete with
    highly efficient time evolution algorithms, boundary conditions,
    diagnostics, and checkpointing capabilities.}
  \label{tab:SENR_NRPy}
  \begin{tabular}{|p{0.12\linewidth}|p{0.76\linewidth}|l|}
    \hline
    \hline
    Task & Description & Section \\
    \hline
    \hline
    Coordinates & The curvilinear coordinates
    are defined in terms of the uniform coordinates within
    \textbf{NRPy+}. Only the scale factors (the square root of each
    diagonal reference-metric component) must be defined; all hatted
    quantities in Eq.~\eqref{eq:BSSN} are evaluated directly from
    these scale factors. In addition, mappings from the chosen
    curvilinear coordinate basis to spherical and 
    Cartesian coordinate bases are defined. The former is
    necessary for transforming initial data (currently expressed
    exclusively in spherical coordinates) to the desired
    uniform coordinates. The latter is necessary to transform from the
    chosen basis to evaluate the ADM integrals (expressed in Cartesian
    coordinates for convenience in interpreting the linear and angular
    momentum components).
    &
    \ref{sec:coordinate_options} \\
    \hline
    Initial Data & Various initial configurations are
    included inside \textbf{NRPy+}, including multiple black hole
    Brill-Lindquist~\cite{Brill:1963} initial data, conformally 
    curved UIUC~\cite{Liu:2009} initial data for single Kerr black holes, as
    well as analytical static trumpet initial data~\cite{Dennison:2014}. In
    addition, there are several choices for the initial gauge 
    conditions. As described above, all initial data are currently
    written in a spherical basis, and converted to the desired
    curvilinear coordinate basis by \textbf{NRPy+} . \textbf{SENR}
    reads the initial data code generated by \textbf{NRPy+} to define each
    of the $24$ BSSN fields evolved on our grids (``grid functions'') at the 
    initial time. & 
    \ref{sec:initial_data} \\
    \hline
    Boundary \quad\quad Conditions & 
    \textbf{SENR} fills the ghost zone points with data from
    the grid interior for each of the evolved grid functions. Specialized
    boundary condition routines are written by hand for each
    type of boundary condition, whether they be inner (e.g.,
    $\theta<0$ in spherical-like coordinates) or outer boundary
    conditions. &
    \ref{sec:boundary_conditions} \\ 
    \hline
    Finite \quad\quad\quad\quad\quad Differencing & Where spatial
    derivatives appear, \textbf{NRPy+} constructs finite difference
    stencils at the user-specified accuracy order on the uniform
    grid. Upwinded derivatives are enabled by default
    on all shift-advection terms (as is typical; see e.g.,
    \cite{Alcubierre:2003b, Thornburg:2004b, Chirvasa:2008}). &
    \ref{sec:numerical_differentiation} \\ 
    \hline
    Evolution \quad Equations & \textbf{NRPy+} constructs and outputs the
    evolution equation right-hand-side C codes, expressing all spatial
    derivatives of the grid functions as finite differences. The C
    codes include reading all needed data from memory. &
    \ref{sec:evolution_equations} \\
    \hline
    Diagnostics & \textbf{NRPy+} outputs the needed C codes for the BSSN
    constraints, the ADM integrands, and the spherically symmetric
    horizon finder. Additionally, \textbf{SENR} includes a code that
    interpolates grid function data onto a Cartesian grid to be
    evaluated using the large suite of diagnostic utilities within the
    ETK, such as generic horizon finding and
    gravitational wave extraction. Diagnostics are
    periodically evaluated by \textbf{SENR} and stored to disk. & 
    \ref{sec:diagnostics} \\
    \hline
    Numerical \quad Grids & \textbf{SENR} allocates memory for coordinate grids and
    evolved grid functions given the number of grid points and the
    coordinate definitions. & 
    \ref{sec:grids} \\
    \hline
    Time \quad\quad\quad Integration & \textbf{SENR} computes the
    largest-allowed CFL time step from the proper distance, determined
    by the reference metric (output from \textbf{NRPy+}) and the chosen
    grid. It iterates the grid functions to the next time step, evaluating
    the evolution equation C codes. Generally, RK4 is used for time
    integrations. & 
    \ref{sec:time_integration} \\
    \hline
    \hline
  \end{tabular}
\end{table*}

\subsection{Computational grid structures}
\label{sec:grids}

Each of the 24 evolved fields defined in Sec.~\ref{sec:evolution_equations}
are sampled by a discrete computational grid, represented as a numerical
array storing the function value at each grid point. We define a uniformly sampled
unit cube grid with coordinate labels
$x^{(i)} = \left(x^{1}, x^{2}, x^{3}\right) \equiv \left(\xOne, \xTwo, \xThree\right)$,
where $x^{1}$ represents the first spatial degree of freedom in Einstein notation and
$\xOne$ represents the first coordinate as it appears in SENR/NRPy+,
and so on for the other two coordinates. These coordinates correspond
to the rescaled tensor basis, in which coordinate singularities have
been removed. Thus we perform numerical integrations and finite difference operations
on this grid using ordinary, uniformly spaced stencils.
The uniform coordinates are mapped to the nonuniformly distributed Cartesian coordinates
$y^{(i)} = y^{(i)}\left(x^{(j)}\right) \equiv \left(\yOne,\yTwo,\yThree\right)$, chosen
to exploit the near-symmetries of the physical system of interest. Tensors in the
$y^{(i)}$ coordinates exist in a Cartesian coordinate basis with trivial symmetry and parity
conditions (i.e., no inner boundaries).

The user specifies the number of grid points $N_{i}$ dedicated to each
coordinate direction $x^{i}$, fixing the uniform grid cell spacing
\beq
\Delta x^{i} = \frac{1}{N_{i}} \; .
\eeq
The current method requires $N_{i}$ to be even and $N_{i} \geq 2$. The
user also chooses the finite difference order $N_{\text{FD}}$ (see
Sec.~\ref{sec:numerical_differentiation}), which determines the number
of ``ghost zone'' points $N_{\text{G}} = N_{\text{FD}} / 2 + 1$ on either side
of the domain required to evaluate finite difference stencils that extend
beyond the boundary. Most derivatives are computed using centered stencils,
extending $N_{\text{FD}} / 2$ grid points symmetrically to either side of the
point in question. Shift-advection derivatives (terms acted on by $\beta^{i} \partial_{i}$)
are approximated by upwinded finite differences, which employ asymmetrical
stencils with  $N_{\text{FD}} / 2 + 1$ points on one side and $N_{\text{FD}} / 2 - 1$
on the other. Points in the ghost zone are not evolved directly, but 
depend entirely on the grid interior and are updated by the boundary condition routine 
(see Sec.~\ref{sec:boundary_conditions}). Thus, the total number of 
points allocated to each coordinate is
\beq
  N_{\text{T} i} = N_{i} + 2 N_{\text{G}} \; .
\eeq
The total number of points on the uniform grid is simply
\beq
  N_{\text{T}} = \prod_{i} N_{\text{T} i} \; .
\eeq
The grid points themselves are located at
\beq
  x^{(i)}(j) = \Delta x^{i} \left (j - N_{\text{G}} + \frac{1}{2} \right ) \; ,
\eeq
where the grid index 
$j \in \left\{0, 1, \ldots, N_{\text{T} i} - 1\right\}$. This produces
a grid that guarantees functions are never evaluated on the
coordinate singularities at, e.g., $r = 0$, $\theta = 0$, or 
$\theta = \pi$ in spherical coordinates. Points for which 
all three coordinates satisfy $0 < x^{i} < 1$ are part of the grid interior, whereas points 
with any of the three $x^{i} < 0$ or $x^{i} > 1$ are in the ghost zone.

\subsubsection{Coordinate options}
\label{sec:coordinate_options}

In the present work, we demonstrate BSSN evolution in three different
classes of coordinate system: Cartesian-like, cylindrical-like, and
spherical-like. These are distinguished by the number and character of
their inner boundary conditions. Boundary conditions must be applied on
all faces of our numerical grid cube, whether the face maps to another
face (e.g., in the case of the $\varphi = 0$ and $\varphi = 2 \pi$ faces in
spherical coordinates) or corresponds to the outer boundary (e.g., at
$r = r_\text{max}$ in spherical coordinates). The details of how
to specify the inner and outer boundary ghost zone points are discussed in
Sec.~\ref{sec:boundary_conditions}.

Finite difference stencils are evaluated on the
uniform $x^{(i)}$ grid, which possesses a
one-to-one mapping to the nonuniformly sampled Cartesian coordinates 
$y^{(i)}$. For example, in spherical-like
coordinates, the nonuniform grid is related to the uniform grid via
\begin{subequations}
\begin{align}
  \yOne &= x = f\left(\xOne\right) \sin\left(\pi \, \xTwo\right) \cos\left(2 \pi \, \xThree\right) \; , \\
  \yTwo &= y = f\left(\xOne\right) \sin\left(\pi \, \xTwo\right) \sin\left(2 \pi \, \xThree\right) \; , \\
  \yThree &= z = f\left(\xOne\right) \cos\left(\pi \, \xTwo\right) \; .
\end{align}
\end{subequations}
The coordinate distribution is generalized by the function $f$ which
is required to be invertible, odd with respect to the origin, and at
least twice differentiable. These properties of $f$ determine
the symmetry conditions for the inner boundaries of the grid, which are described in detail
for spherical-like coordinates and the general case in Sec.~\ref{sec:boundary_conditions}.
The choice $f(\xOne) = r_{\text{max}} \, \xOne$ reduces
this to ordinary, uniform spherical coordinates extending out to
radius $r_{\text{max}} / M > 0$. It is often useful to adopt a
logarithmically distributed radial coordinate of the form 
$f(\xOne) = A \sinh(\xOne / w)$, where $A, w > 0$ are free parameters. This allows the
radial outer boundary to be pushed far away while maintaining high
resolution near the origin. Another possibility is to take 
$f(\xOne) = a \, \xOne + b \, \xOne^{3} + c \, \xOne^{5}$, where appropriate choices 
for the coefficients
$a$, $b$, and $c$ lead to increasing the relative coordinate density
on a spherical shell, which is ideal for sampling a black hole horizon
or neutron star crust.

The rescaling procedure developed in this paper also allows for the
angular coordinates to be redistributed in a similar way, but we
restrict our discussion to radial rescalings for the sake of
simplicity. To be clear, by ``radial,'' we refer to
$\{\xOne,\xTwo,\xThree\}$ in Cartesian-like, to $\{\xOne,\xThree\}$
(cylindrical radius and height) in cylindrical-like, and to $\{\xOne\}$
(radius) in spherical-like coordinates.

To construct the noncoordinate basis for use in the tensor component rescaling,
we start with the flat metric in the $y^{(i)}$ Cartesian coordinate basis. Next,
transform it to the coordinate basis of the uniform grid $x^{(i)}$ using the
Jacobian matrix. We identify $\hat{\gamma}_{i j}$ with the flat metric in the $x^{(i)}$ basis.
Then, the noncoordinate basis components are defined
by Eq.~\eqref{eq:basis_definition}. When the coordinate system is
orthogonal, as is the case for all examples in Sec.~\ref{sec:results}, then 
only three of the nine basis
components are nonzero.

The coordinate system distributions and the corresponding basis
components are summarized in Table~\ref{tab:coordinates}.

\begin{table*}[!t]
  \centering
  \caption{Summary of coordinate system choices. Finite difference
    operations take place on the uniformly sampled unit cube grid
    $x^{(i)} = \left(\xOne, \xTwo, \xThree\right)$, and are mapped to the
    nonuniformly sampled Cartesian grid $y^{(i)} = \left(\yOne, \yTwo, \yThree\right)$. The
    nontrivial scale factors constitute the basis, which is used to
    rescale tensor quantities. The function $f$ allows the coordinates
    to be redistributed on the nonuniform grid, and the prime mark
    indicates differentiation with respect to the function
    argument. Although not shown here, our method allows for
    the more general case of a different redistribution function for
    each independent coordinate.}
  \label{tab:coordinates}
  \begin{tabular}{| l | c | c | c | c | c | c |}
    \hline
    \hline
    Coordinates & \multicolumn{3}{|c|}{Definitions} & \multicolumn{3}{|c|}{Scale Factors} \\
    \hline
    \hline
     & $\yOne$ & $\yTwo$ & $\yThree$ & $e_{\xOne}^{(\xOne)}$ & $e_{\xTwo}^{(\xTwo)}$ & $e_{\xThree}^{(\xThree)}$ \\
    \hline
    Cartesian-like & $f\left(\xOne\right)$ & $f\left(\xTwo\right)$ & $f\left(\xThree\right)$ & $f'\left(\xOne\right)$ & $f'\left(\xTwo\right)$ & $f'\left(\xThree\right)$ \\
    Cylindrical-like & $f\left(\xOne\right) \cos\left(2 \pi \, \xTwo\right)$ & $f\left(\xOne\right) \sin\left(2 \pi \, \xTwo\right)$ & $f\left(\xThree\right)$ & $f'\left(\xOne\right)$ & $f\left(\xOne\right)$ & $f'\left(\xThree\right)$ \\
    Spherical-like & $f\left(\xOne\right) \sin\left(\pi \, \xTwo\right) \cos\left(2 \pi \, \xThree\right)$ & $f\left(\xOne\right) \sin\left(\pi \, \xTwo\right) \sin\left(2 \pi \, \xThree\right)$ & $f\left(\xOne\right) \cos\left(\pi \, \xTwo\right)$ & $f'\left(\xOne\right)$ & $f\left(\xOne\right)$ & $f\left(\xOne\right) \sin\left(\pi \, \xTwo\right)$ \\
    \hline
    \hline
  \end{tabular}
\end{table*}

\subsubsection{Numerical representation of spatial derivatives}
\label{sec:numerical_differentiation}

The user-specified $n = N_{\text{FD}}$ sets the order of the finite
difference stencil approximation, so that the truncation error scales
as 
\beq
\mathcal{E}_{\text{FD}} \sim \mathcal{O}\left(\Delta x^{n} \left|\partial_{x}^{n+1} u\right| \right)
\eeq
for each of the 24 dynamical fields 
${u = \{\varepsilon_{i j}, \bar{A}_{i j}, W, K, \bar{\Lambda}^{i}, \alpha, \beta^i, B^{i}\}}$.
We demonstrate that finite-difference truncation errors
converge to zero at the prescribed rates in Sec.~\ref{sec:grid_convergence}.
To calculate the finite difference stencil coefficients, NRPy+ inverts
the corresponding linear system of Taylor series
coefficients at user-specified order, akin to inverting the
Vandermonde matrix for Lagrange polynomial interpolation~\cite{Macon:1958}. Adopting a
simple syntax, NRPy+ automatically replaces all spatial derivatives
that appear in expressions with the appropriate finite difference
approximation, at the desired order.
Such spatial derivatives appear throughout the right-hand sides
of the evolution system, Eqs.~\eqref{eq:BSSN}, \eqref{eq:1_plus_log},
and~\eqref{eq:Gamma-driver}, and diagnostics---including the BSSN
constraint equations and ADM integrals (see Sec.~\ref{sec:diagnostics}). 

We use Kreiss-Oliger dissipation~\cite{Kreiss:1973, Alcubierre:2008}
to diffuse unresolved, high-frequency modes that can reduce
the convergence order. A standard high-order
derivative operator $\mathcal{L}_{\text{KO}}$ acting on the grid function as
\beq
  \mathcal{L}_{\text{KO}} u = -\epsilon_{\text{KO}} \frac{(-1)^{n/2}}{2^{n} \Delta t} \left(\Delta x^{i} \right)^{n} \partial_{i}^{n} u
\eeq
is added to the right-hand sides of the evolution equations~\eqref{eq:BSSN},
\eqref{eq:1_plus_log}, and~\eqref{eq:Gamma-driver}. 
Note that $\mathcal{L}_{\text{KO}}$ is not a tensorial derivative in general, and its
inclusion in the evolution equations violates spatial
covariance. However, the coefficient of artificial dissipation is
chosen such that the contribution of $\mathcal{L}_{\text{KO}} u$ vanishes in the
continuum limit. The dimensionless Kreiss-Oliger parameter
$\epsilon_{\text{KO}} = \epsilon_{\text{KO}}\left(x^{i}\right)$ is allowed to vary smoothly
over space, and typically approaches
$\epsilon_{\text{KO}} = 0.99$ in the weak field
region. In particular, we often use a spherically symmetric transition
function 
\beq
  \epsilon_{\text{KO}}(r) = \frac{\epsilon_{\text{KO0}}}{2} \left [\mathrm{erf}\left(\frac{r - r_{\text{KO}}}{w_{\text{KO}}}\right) + 1 \right ] \; ,
\eeq
where $\epsilon_{\text{KO0}}, r_{\text{KO}}, w_{\text{KO}} > 0$ are constant
parameters and $r$ is a radial coordinate. We generally set this function to be less than $10^{-16}$
near the origin, so that its nonsmoothness at the origin is made
irrelevant relative to nonsmoothness caused by roundoff error.
In particular, we usually set ${\epsilon_{\text{KO0}} = 0.99}$,
${r_{\text{KO}} / M = 2}$, and ${w_{\text{KO}} / M = 0.17}$.

\subsection{Diagnostics}
\label{sec:diagnostics}

We employ a variety of diagnostics to monitor the
accuracy of our calculations, as well as to probe the physical
properties of the simulated spacetimes. Diagnostic routines fall broadly into two
categories: diagnostics generated in NRPy+, and
diagnostic routines within the ETK. 

\subsubsection{Diagnostics generated by NRPy+}
\label{sec:NRPy_diagnostics}

The following describes the constraints, the
ADM integrals, and a spherically symmetric
horizon finder, which are the diagnostics written in Python
in NRPy+. They contain spatial derivatives of the evolved fields,
which are approximated by the automatically generated finite
difference stencils. The resulting C code is evaluated by SENR during
data output, after the time step iteration.

In terms of the BSSN variables (see
Sec~\ref{sec:evolution_equations}), the Hamiltonian constraint takes
the form~\cite{Baumgarte:2013}
\begin{align}
  \label{eq:Hamiltonian}
  \mathcal{H} &\equiv \frac{2}{3} K^{2} - \bar{A}_{i j} \bar{A}^{i j} + e^{-4 \phi} \left (\bar{R} - 8 \bar{D}^{i} \phi \bar{D}_{i} \phi - 8 \bar{D}^{2} \phi \right ) \nonumber \\
  &= 0 \; ,
\end{align}
where $\bar{R} = \bar{\gamma}^{i j} \bar{R}_{i j}$, and the momentum constraint is\footnote{The term $\bar{A}^{i k} \Delta_{j k}^{j}$ in Eq.~\eqref{eq:Momentum} of this paper is missing from the momentum constraint in Eq.~$(17)$ of~\cite{Brown:2009} and Eq.~$(14)$ of~\cite{Baumgarte:2013}. In the notation of their respective articles, the expression in~\cite{Brown:2009} is corrected by the substitution $g \to g / \tilde{g}$, and in~\cite{Baumgarte:2013} by $\bar{\gamma} \to \bar{\gamma} / \hat{\gamma}$. Be mindful of a parenthesis size mismatch in Eq.~$(14)$ of~\cite{Baumgarte:2013}.}
\begin{align}
  \label{eq:Momentum}
  \mathcal{M}^{i} &\equiv e^{-4 \phi} \left (\hat{D}_{j} \bar{A}^{i j} + 2 \bar{A}^{k (i} \Delta^{j)}_{j k} + 6 \bar{A}^{i j} \partial_{j} \phi - \frac{2}{3} \bar{\gamma}^{i j} \partial_{j} K \right ) \nonumber \\
  &= 0 \; .
\end{align}
The Hamiltonian, momentum, and conformal connection coefficient~\eqref{eq:conformal_connection_constraint}
constraints are monitored
throughout the simulation as a measure of numerical accuracy. In
addition, the ADM surface integrals for total mass $M_{\text{ADM}}$,
linear momentum $P_{\text{ADM}}^{i}$, and angular momentum
$J_{\text{ADM}}^{i}$ also serve as diagnostics. The integrands are
evaluated on a spherical surface on the boundary of the spatial
hypersurface at spatial infinity. Numerically, the integrals are
approximated by two-dimensional Riemann sums on a spherical surface
which is near the outer boundary, ideally in the weak field
region. Supposing that the spacetime is asymptotically flat, and that
the spacetime metric $g_{\mu \nu}$ approaches the Minkowski metric
$\eta_{\mu \nu}$ at least as fast as $g_{\mu \nu} - \eta_{\mu \nu} =
\mathcal{O}(1/r)$ when $r \to \infty$, then the ADM integrals take the
form~\cite{Baumgarte:2010} 
\begin{subequations}
  \label{eq:ADM_Integrals}
\begin{align}
  \label{eq:ADM_Mass}
  M_{\text{ADM}} &= \lim_{r \to \infty} \frac{1}{16 \pi} \oint \gamma^{i j} \left (\partial_{i} \gamma_{k j} - \partial_{k} \gamma_{i j} \right ) \sqrt{\gamma} \, \ud S^{k} \; , \\
  \label{eq:ADM_Momentum}
  P_{\text{ADM}}^{i} &= \lim_{r \to \infty} \frac{1}{8 \pi} \oint \left (K^{i j} - \gamma^{i j} K \right) \sqrt{\gamma} \, \ud S_{j} \; , \\
  \label{eq:ADM_Angular_Momentum}
  J_{\text{ADM}}^{i} &= \lim_{r \to \infty} \frac{[i j k]}{8 \pi} \oint y_{j} \left (K_{k l} - \gamma_{k l} K \right) \sqrt{\gamma} \, \ud S^{l} \; .
\end{align}
\end{subequations}
The vector components $\ud S^{i}$ play the role of the
outward-oriented surface element induced at spatial infinity, $y^{i}$
are the components of a Cartesian coordinate vector, and $[i j k]$ is
the totally antisymmetric Levi-Civita symbol. Note that the ADM
integrals are not covariant as written, and they must be evaluated in
asymptotically Cartesian coordinates. This allows us to easily interpret the
directionality of the $P_{\text{ADM}}^{i}$ and $J_{\text{ADM}}^{i}$ components.

In the special case of a spherically symmetric configuration, the
expansion of outgoing null geodesics $\Theta$ takes the simplified
form~\cite{Thornburg:1999}
\beq
  \label{eq:null_expansion}
  \Theta(r) = \frac{4 \bar{\gamma}_{\theta \theta} \partial_{r} \phi + \partial_{r} \bar{\gamma}_{\theta \theta}}{e^{2 \phi} \bar{\gamma}_{\theta \theta} \sqrt{\bar{\gamma}_{r r}}} - 2 \frac{\bar{K}_{\theta \theta}}{\bar{\gamma}_{\theta \theta}} \; .
\eeq
The coordinate radius of the apparent horizon $r_{\text{H}}$ is
defined to satisfy $\Theta(r_{\text{H}}) = 0$. Numerically, SENR
evaluates $\Theta$ at every grid point using finite difference stencil
C code generated by NRPy+. Then, it searches for the pair of neighbors
that straddle $\Theta = 0$, and linearly interpolates between those
two points to approximate $r_{\text{H}}$.

\subsubsection{Diagnostics provided by the Einstein Toolkit}

A translation layer for the ETK is implemented in SENR,
where the fields $\bar{\gamma}_{i j}$, $e^{\phi}$, $\bar{A}_{i j}$,
and $K$ are interpolated onto a Cartesian grid and then converted to
the ADM quantities $\gamma_{ij}$ and $K_{ij}$ in the Cartesian
basis. These data are fed into the ETK to unlock a
wide variety of diagnostic tools~\cite{Dreyer:2003, Szabados:2004, Schnetter:2006}, including the horizon finder thorn
AHFinderDirect~\cite{Thornburg:2004a} and the $\Psi_{4}$ gravitational
waveform extraction thorn WeylScal4~\cite{Baker:2002,
  Pollney:2009}. The measured $\Psi_{4}$ contains information relating to the gravitational wave
strain in the transverse-traceless gauge and the weak field
region via~\cite{Baumgarte:2010} 
\beq
  \label{eq:Weyl_scalar}
  \Psi_{4} = \ddot{h}_{+} - i \ddot{h}_{\times} \; ,
\eeq
where $h_{+}$ and $h_{\times}$ are the gravitational wave strain
amplitudes of the ``plus'' and ``cross'' polarization states,
respectively, and the dots denote time derivatives. 

\subsection{Initial data}
\label{sec:initial_data}

NRPy+ implements initial data for zero- ($\gamma_{i j} = \hat{\gamma}_{i j}$),
one-, and two-black-hole spacetimes. Single Kerr black hole initial data are available in
UIUC conformally curved coordinates~\cite{Liu:2009}, Schwarzschild trumpet
coordinates~\cite{Dennison:2014}, and boosted Schwarzschild black
holes in isotropic coordinates~\cite{Ruchlin:2017}. Two-black-hole 
initial data take the form of initial black holes at rest
(Brill-Lindquist~\cite{Brill:1963}). All implemented initial data
solve the Hamiltonian~\eqref{eq:Hamiltonian} and
momentum~\eqref{eq:Momentum} constraints exactly. Expressions for all
initial data evolved in this work are presented alongside their
results in Sec.~\ref{sec:results}. 

Typically, these initial data types are most naturally represented in
the spherical coordinate basis. The Jacobian matrix is used to
transform the initial data from uniform spherical coordinates to the
desired uniformly sampled $x^{(i)}$ grid. Finally, rank-1 and -2 tensors are
transformed to the noncoordinate basis, according to the procedure in
Sec.~\ref{sec:tensor_rescaling}.

\subsection{Time integration}
\label{sec:time_integration}

The evolution equations~\eqref{eq:BSSN}, \eqref{eq:1_plus_log},
and~\eqref{eq:Gamma-driver} are all first-order-in-time partial
differential equations that may be written in the form
\beq
  \partial_{t} u(t) = \mathcal{L}\left(u(t), t\right) \; ,
\eeq
where 
$u = \{\varepsilon_{i j}, \bar{A}_{i j}, W, K, \bar{\Lambda}^{i}, \alpha, \beta^i, B^{i}\}$ 
is a vector composed of the 24 evolved fields. As can be seen from
Eqs.~\eqref{eq:BSSN}, \eqref{eq:1_plus_log},
and~\eqref{eq:Gamma-driver}, the differential operator $\mathcal{L}$
depends on multiple components of $u$, as well as
their first and second spatial derivatives. All spatial derivatives of
the evolved fields in $\mathcal{L}$ are calculated using finite
differences on the uniformly sampled $x^{(i)}$ grid. 

To advance the grid function in time ${u(t) \to u(t + \Delta t)}$, we
adopt the fourth-order\footnote{See Ref.~\cite{private_comm}.
See also~\cite{Cordero-Carrion:2012} for details on the stability properties
of the fully explicit Runge-Kutta methods at various order. The widely
used RK4 method is conditionally stable for this application, although PIRK4
might allow for similar accuracy with larger timesteps.} Runge-Kutta (RK4) method~\cite{Press:2007}
\begin{subequations}
\begin{align}
  k_{1} &= \mathcal{L}\left(u(t), t\right) \; , \\
  \label{eq:RK4_sub1}
  u_{1} &= u(t) + \frac{\Delta t}{2} k_{1} \; , \\
  k_{2} &= \mathcal{L}\left(u_{1}, t + \frac{\Delta t}{2} \right ) \; , \\
  \label{eq:RK4_sub2}
  u_{2} &= u(t) + \frac{\Delta t}{2} k_{2} \; , \\
  k_{3} &= \mathcal{L}\left(u_{2}, t + \frac{\Delta t}{2} \right ) \; , \\
  \label{eq:RK4_sub3}
  u_{3} &= u(t) + \Delta t k_{3} \; , \\
  k_{4} &= \mathcal{L}\left(u_{3}, t + \Delta t \right ) \; , \\
  u(t + \Delta t) &= u(t) + \frac{\Delta t}{6} \left (k_{1} + 2 k_{2} + 2 k_{3} + k_{4} \right ) \; .
\end{align}
\end{subequations}
Being fourth order means that the error associated with the time
stepping, at a fixed time, scales as 
${\mathcal{E}_{\text{RK4}} \sim \mathcal{O}\left(\Delta t^{4}\right)}$.
Immediately after evaluating the RK4 steps given by
Eqs.~\eqref{eq:RK4_sub1},~\eqref{eq:RK4_sub2}, and~\eqref{eq:RK4_sub3},
boundary conditions are applied to $u_{1}$, $u_{2}$, and $u_{3}$,
respectively. If the boundary conditions are time dependent, then they
are applied at $t + \Delta t / 2$ on the first two substeps and at a
full step $t + \Delta t$ on the third substep. At the end of the full
RK4 iteration, boundary conditions are applied to $u$ at time 
$t + \Delta t$. Finally, the algebraic correction
\beq
  \bar{\gamma}_{i j} \to \left (\frac{\hat{\gamma}}{\bar{\gamma}} \right )^{1/3} \bar{\gamma}_{i j}
\eeq
is applied to the metric components to enforce the Lagrangian
specification constraint~\eqref{eq:Lagrangian_detg}, where
$\bar{\gamma} = \hat{\gamma}$ is the conformal metric determinant on the initial
slice.

When applying this standard, explicit RK4 algorithm, the
Courant-Friedrichs-Lewy (CFL) condition~\cite{Press:2007} must be
satisfied. For a numerical grid with coordinates $x^{(i)}$, SENR finds
the smallest proper distance $\Delta s_{\min}$ along each of the
independent coordinate directions
\beq
  \Delta s_{\min} = \min\left(\sqrt{\bar{\gamma}_{1 1}} \Delta \xOne, \sqrt{\bar{\gamma}_{2 2}} \Delta \xTwo, \sqrt{\bar{\gamma}_{3 3}} \Delta \xThree \right) \; ,
\eeq
where $\Delta x^{i}$ is the uniform grid spacing between adjacent
points in the $x^{i}$ direction, and $\bar{\gamma}_{i i}$ is evaluated
at $x^{i}$. Then the time step is
\beq
  \Delta t = C \Delta s_{\text{min}} \; ,
\eeq
where the Courant factor is set to $C = 0.5$ for all simulations
presented here.

More explicitly, the CFL-limited time step varies with grid
resolution in a nontrivial way depending on the coordinate choice. Suppose that we
vary the number of grid points simultaneously in all three
coordinates. Then for Cartesian coordinates $\Delta t \propto \Delta
x^{i}$, for cylindrical coordinates $\Delta t \propto \left(\Delta
x^{i}\right)^{2}$, and for spherical coordinates $\Delta t \propto
\left(\Delta x^{i}\right)^{3}$. The higher-order dependence in the
case of cylindrical and spherical coordinates is due to the focusing
of grid points along the symmetry axis or near the
origin. This CFL restriction can be softened significantly by clever
choice of coordinate redistribution function $f$ (see
Sec.~\ref{sec:coordinate_options}). 

\subsection{Boundary conditions}
\label{sec:boundary_conditions}

As described in Sec.~\ref{sec:grids}, SENR/NRPy+ maps a uniformly
sampled unit cube grid to a nonuniformly distributed Cartesian coordinate grid, chosen to
efficiently sample the space. Our grids are cell centered, so no
points exist precisely on any of the faces of the unit cube. However, for the points
that are nearest to the faces, but inside the cube, the finite difference
stencils for spatial derivatives will reach outside of the cube. To
ensure that these stencils correspond to valid data, we add a
collection of points to a shell region exterior to the cube called the ``ghost zone.''
If $N_{\text{G}}$ ghost zone points are needed
outside each boundary, then this would increase the total number of
points in the grid from 
$N_{\xOne} \times N_{\xTwo} \times N_{\xThree}$ to 
$(N_{\xOne}+2 N_{\text{G}}) \times (N_{\xTwo}+2 N_{\text{G}}) \times (N_{\xThree}+2 N_{\text{G}})$.

Prior to evaluation of the right-hand sides of the BSSN equations,
these ghost zone points must be filled. Some ghost zone points, including those at
$\varphi < 0$ or $\varphi > 2 \pi$ on spherical- or cylindrical-like coordinate
grids, map to points {\it inside} the cube; we call these {\it inner}
boundaries. The remaining ghost zone points map back to points outside the
interior of the cube (e.g., $r > r_{\rm max}$ in spherical-like
coordinate grids); we call these {\it outer} boundaries.

Outer boundary ghost zone points may be filled in accordance with the
desired outer boundary condition. Although the widely used Sommerfeld
outer boundary condition is also implemented, we find the simple
quadratic extrapolation condition to be quite effective on
spherical-like coordinate grids
\beq
  \label{eq:quadratic_extrapolation}
  u(x) \approx 3 u(x - \Delta x) - 3 u(x - 2 \Delta x) + u(x - 3 \Delta x) \; ,
\eeq
where $u$ is any of the evolved fields and $x$ is the coordinate $x^{(i)}$ perpendicular to the boundary.
As with any approximate boundary condition, this condition
produces unwanted ingoing modes that contaminate the interior of the
simulation. In practice, logarithmically spaced radial
coordinates enable us to push the outer boundary out of causal contact
with the origin for as long as we care to simulate.

Inner boundary conditions depend on the coordinate system,
and must account for intrinsic periodic, axial, and radial
symmetries. The coordinate redistribution function $f$ (see
Sec.~\ref{sec:coordinate_options}) is required to be odd, which ensures
that ghost zone points across inner boundaries coincide exactly with
other points on the grid interior, respecting the desired symmetries.

In the case of scalar functions, these symmetry conditions simply copy
the appropriate values of the function from the grid interior to its
ghost zone partner. Vectors and higher rank tensors, however, are sensitive
to changes of sign in the basis when evaluated across inner
boundaries. In that case, an appropriate change of sign must be copied
into the ghost zone along with the function value itself. We refer to
these changes of sign as {\it parity conditions}.

In the following, we will show by example the symmetry and parity
conditions specific to the spherical coordinate topology. Then a
generic algorithm for assigning ghost zone values in arbitrary
coordinates is described.

\subsubsection{Spherical boundary conditions example}

To derive the boundary conditions appropriate for a coordinate system,
first express those coordinates in terms of a system whose boundaries
are well understood. To this end, we choose ordinary Cartesian
coordinates on the domain $-\infty < x, y, z < \infty$. Since each
coordinate is unbounded from both above and below, there are
no inner boundaries. In addition, every point on the computational
grid has a unique $(x, y, z)$ label.

Now consider ordinary spherical coordinates, which are
related to the Cartesian coordinates in the usual
way [Eq.~\eqref{eq:Cartesian_spherical}]. The spherical coordinate domain is
bounded by $0 < r < \infty$, $0 < \theta < \pi$, and $0 < \varphi < 2
\pi$. In this case, there is only one outer boundary, corresponding to
$r \to \infty$; what remains are the five inner boundaries. 

To find the symmetry conditions for these inner boundaries, first
recall that a scalar function $g$ has a particular value at some
location, regardless of the underlying coordinate choice. Next, evaluate the function in
the spherical coordinate ghost zone, identify the corresponding
Cartesian coordinate values, and then find the point in the spherical
grid interior that corresponds to those same Cartesian
coordinates. This links each point in the ghost zone to
a unique point in the grid interior.

By this procedure, the five symmetry conditions for the spherical
coordinate inner boundaries are found to be 
\begin{subequations}
  \label{eq:spherical_symmetry_conditions}
  \begin{align}
    \label{eq:radial_symmetry}
    g(-r, \theta, \varphi) &= g(r, \pi - \theta, \pi + \varphi) \; , \\
    \label{eq:axial_N_symmetry}
    g(r, -\theta, \varphi) &= g(r, \theta, \pi + \varphi) \; , \\
    \label{eq:axial_S_symmetry}
    g(r, \pi + \theta, \varphi) &= g(r, \pi - \theta, \pi + \varphi) \; , \\
    \label{eq:periodic_n_symmetry}
    g(r, \theta, -\varphi) &= g(r, \theta, 2 \pi - \varphi) \; , \\
    \label{eq:periodic_p_symmetry}
    g(r, \theta, 2 \pi + \varphi) &= g(r, \theta, \varphi) \; .
\end{align}
\end{subequations}
These correspond to radial symmetry about the origin [Eq.~\eqref{eq:radial_symmetry}],
axial symmetry about the north [Eq.~\eqref{eq:axial_N_symmetry}] and south
[Eq.~\eqref{eq:axial_S_symmetry}] poles, and periodic symmetry around the axis in
the negative [Eq.~\eqref{eq:periodic_n_symmetry}] and positive
[Eq.~\eqref{eq:periodic_p_symmetry}] orientations.
For any point in the ghost zone, there exists a combination of the inner boundary
symmetry rules~\eqref{eq:spherical_symmetry_conditions} that maps that point
to either the grid interior or the outer boundary. Note that these symmetries
refer only to the coordinate distribution, and not
the evolved fields, which are allowed to be completely asymmetrical.

As mentioned above, these symmetry conditions are sufficient for
filling the inner boundary ghost zone points of a scalar function. For
the case of vectors and higher rank tensors, however, parity conditions
must also be taken into account. 

The needed parity conditions are found by comparing the basis vectors
in the ghost zone to their counterparts in the grid interior. Again,
we express the basis vectors in terms of a basis in which the parity
conditions are well understood. Since the Cartesian basis has no inner
boundaries, the Cartesian basis has only the trivial parity
conditions, which is to say that there are no changes of sign.
Start with the noncoordinate spherical basis
$\boldsymbol{e}_{(a)}$~\eqref{eq:spherical_vector_basis} which 
is expressed in terms of the spherical coordinate basis
$\frac{\boldsymbol{\partial}}{\boldsymbol{\partial} r^{(a)}}$. Then,
transform the spherical coordinate basis to the Cartesian
basis~\eqref{eq:chain_rule}
\begin{subequations}
\begin{align}
  \boldsymbol{e}_{(r)} &= \frac{\partial x^{i}}{\partial r} \frac{\boldsymbol{\partial}}{\boldsymbol{\partial} x^{(i)}} \; , \\
  \boldsymbol{e}_{(\theta)} &= \frac{1}{r} \frac{\partial x^{i}}{\partial \theta} \frac{\boldsymbol{\partial}}{\boldsymbol{\partial} x^{(i)}} \; , \\
  \boldsymbol{e}_{(\varphi)} &= \frac{1}{r \sin(\theta)} \frac{\partial x^{i}}{\partial \varphi} \frac{\boldsymbol{\partial}}{\boldsymbol{\partial} x^{(i)}} \; .
\end{align}
\end{subequations}
Remembering $\boldsymbol{e}_{(i)} =
\frac{\boldsymbol{\partial}}{\boldsymbol{\partial} x^{(i)}}$ in
Cartesian coordinates and contracting with the inverse of the Jacobian
matrix~\eqref{eq:spherical_jacobian} results in
\begin{subequations}
\begin{align}
  \boldsymbol{e}_{(r)}(r, \theta, \varphi) = {} & \sin(\theta) \cos(\varphi) \boldsymbol{e}_{(x)} + \sin(\theta) \sin(\varphi) \boldsymbol{e}_{(y)} \nonumber \\
  & + \cos(\theta) \boldsymbol{e}_{(z)} \; , \\
  \boldsymbol{e}_{(\theta)}(r, \theta, \varphi) = {} & \cos(\theta) \cos(\varphi) \boldsymbol{e}_{(x)} + \cos(\theta) \sin(\varphi) \boldsymbol{e}_{(y)} \nonumber \\
  & - \sin(\theta) \boldsymbol{e}_{(z)} \; , \\
  \boldsymbol{e}_{(\varphi)}(r, \theta, \varphi) = {} & -\sin(\varphi) \boldsymbol{e}_{(x)} + \cos(\varphi) \boldsymbol{e}_{(y)} \; ,
\end{align}
\end{subequations}
where we indicate on the left-hand side the explicit functional
dependence of the spherical noncoordinate basis vectors on
$(r,\theta, \varphi)$. To find the parity conditions, compute the dot
product between a basis vector in the ghost zone and its partner in
the grid interior, according
to Eq.~\eqref{eq:spherical_symmetry_conditions}. If the coordinate system
is properly constructed, then this dot product should always evaluate
to $\pm 1$, where the negative case indicates that the basis changes
sign in the ghost zone. 

For the current example, we find
\begin{subequations}
\begin{align}
  \boldsymbol{e}_{(r)}(-r, \theta, \varphi) \cdot \boldsymbol{e}_{(r)}(r, \pi - \theta, \pi + \varphi) &= -1 \; , \\
  \boldsymbol{e}_{(\theta)}(-r, \theta, \varphi) \cdot \boldsymbol{e}_{(\theta)}(r, \pi - \theta, \pi + \varphi) &= +1 \; , \\
\boldsymbol{e}_{(\varphi)}(-r, \theta, \varphi) \cdot \boldsymbol{e}_{(\varphi)}(r, \pi - \theta, \pi + \varphi) &= -1 \; , \\
  \boldsymbol{e}_{(r)}(r, -\theta, \varphi) \cdot \boldsymbol{e}_{(r)}(r, \theta, \pi + \varphi) &= +1 \; , \\
  \boldsymbol{e}_{(\theta)}(r, -\theta, \varphi) \cdot \boldsymbol{e}_{(\theta)}(r, \theta, \pi + \varphi) &= -1 \; , \\
  \boldsymbol{e}_{(\varphi)}(r, -\theta, \varphi) \cdot \boldsymbol{e}_{(\varphi)}(r, \theta, \pi + \varphi) &= -1 \; , \\
  \boldsymbol{e}_{(r)}(r, \pi + \theta, \varphi) \cdot \boldsymbol{e}_{(r)}(r, \pi - \theta, \pi + \varphi) &= +1 \; , \\
  \boldsymbol{e}_{(\theta)}(r, \pi + \theta, \varphi) \cdot \boldsymbol{e}_{(\theta)}(r, \pi - \theta, \pi + \varphi) &= -1 \; , \\
  \boldsymbol{e}_{(\varphi)}(r, \pi + \theta, \varphi) \cdot \boldsymbol{e}_{(\varphi)}(r, \pi - \theta, \pi + \varphi) &= -1 \; , \\
  \boldsymbol{e}_{(r)}(r, \theta, -\varphi) \cdot \boldsymbol{e}_{(r)}(r, \theta, 2 \pi - \varphi) &= +1 \; , \\
  \boldsymbol{e}_{(\theta)}(r, \theta, -\varphi) \cdot \boldsymbol{e}_{(\theta)}(r, \theta, 2 \pi - \varphi) &= +1 \; , \\
  \boldsymbol{e}_{(\varphi)}(r, \theta, -\varphi) \cdot \boldsymbol{e}_{(\varphi)}(r, \theta, 2 \pi - \varphi) &= +1 \; , \\
  \boldsymbol{e}_{(r)}(r, \theta, 2 \pi + \varphi) \cdot \boldsymbol{e}_{(r)}(r, \theta, \varphi) &= +1 \; , \\
  \boldsymbol{e}_{(\theta)}(r, \theta, 2 \pi + \varphi) \cdot \boldsymbol{e}_{(\theta)}(r, \theta, \varphi) &= +1 \; , \\
  \boldsymbol{e}_{(\varphi)}(r, \theta, 2 \pi + \varphi) \cdot \boldsymbol{e}_{(\varphi)}(r, \theta, \varphi) &= +1 \; .
\end{align}
\end{subequations}

The same procedure can be used to determine the symmetry and parity
conditions for cylindrical-like coordinates. The parity conditions for
cylindrical- and spherical-like coordinates across each inner boundary
are summarized in Table~\ref{tab:parity}.

\begin{table}[!t]
  \centering
  \caption{Cylindrical- and spherical-like parity conditions for rank-1 and rank-2 tensors. ``Radial'' refers to the parity across $r = 0$, ``Axial'' refers to the parity across $\rho = 0$ or $\sin(\theta) = 0$, and ``Periodic'' refers to the parity across $\varphi = 0$ or $\varphi = 2 \pi$. Compare with Table I in~\cite{Baumgarte:2013}.}
  \label{tab:parity}
  \begin{tabular}{|l|c|c|c|c|}
    \hline
    \hline
    Coordinates & Component & Radial & Axial & Periodic \\
    \hline
    \multirow{9}{*}{Cylindrical-like} & $\rho$ & & $-$ & $+$ \\
    & $\varphi$ & & $-$ & $+$ \\
    & $z$ & & $+$ & $+$ \\
    & $\rho \rho$ & & $+$ & $+$ \\
    & $\rho \varphi$ & & $+$ & $+$ \\
    & $\rho z$ & & $-$ & $+$ \\
    & $\varphi \varphi$ & & $+$ & $+$ \\
    & $\varphi z$ & & $-$ & $+$ \\
    & $z z$ & & $+$ & $+$ \\
    \hline
    \multirow{9}{*}{Spherical-like} & $r$ & $-$ & $+$ & $+$ \\
    & $\theta$ & $+$ & $-$ & $+$ \\
    & $\varphi$ & $-$ & $-$ & $+$ \\
    & $r r$ & $+$ & $+$ & $+$ \\
    & $r \theta$ & $-$ & $-$ & $+$ \\
    & $r \varphi$ & $+$ & $-$ & $+$ \\
    & $\theta \theta$ & $+$ & $+$ & $+$ \\
    & $\theta \varphi$ & $-$ & $+$ & $+$ \\
    & $\varphi \varphi$ & $+$ & $+$ & $+$ \\
    \hline
    \hline
  \end{tabular}
\end{table}

Operationally, vector and tensor component values are copied to the
ghost zone using the symmetry condition. For every basis vector
attached to each tensor component, compensate for any changes in sign in the
basis by applying the parity condition.

\subsubsection{The general boundary condition procedure}

The procedure described in the previous section is generalized to
construct an explicit routine for filling inner and outer boundary ghost zone
points in other coordinate systems. Here we review the routine, which
was developed for SENR/NRPy+ to validate the special, coordinate-specific inner and outer
boundary condition routines developed for cylindrical- and
spherical-like grids.

Starting with the mapping between Cartesian coordinates and the
coordinates of interest [e.g., Eq.~\eqref{eq:Cartesian_spherical}],
the routine automatically classifies the ghost zones as inner or outer
boundaries, maps each inner ghost zone point to its partner on the grid
interior, constructs the noncoordinate basis vectors, and determines
the parity conditions. The symmetry and parity conditions are compiled
into a list at the start of a simulation, and the routine later runs
through the list to apply the boundary conditions as needed.

The ghost zone points are filled, in order, outward from the interior
grid. This is of particular importance at the outer boundary, where
the one-sided quadratic extrapolation
stencil equation~\eqref{eq:quadratic_extrapolation} requires that
outer boundary ghost zone points be filled in an outward-going
direction. In addition, for cases of high symmetry and large
$N_{\text{FD}}$, it is possible for the finite difference stencil to
be wider than the grid interior, which means that a ghost zone point
could be mapped to another ghost zone point. Assigning the boundary
values in an outward fashion avoids attempts to copy to the ghost zone
from uninitialized memory.

\section{Results}
\label{sec:results}

In this section we present a number of validation and verification tests performed
on the SENR/NRPy+ code. In Sec.~\ref{sec:code_comparisons}, we compare
its results with two other numerical relativity codes, and demonstrate
that, e.g., differences in results between SENR/NRPy+ and the code of
Baumgarte \emph{et al.}~\cite{Baumgarte:2013} are at the level of roundoff
error in the case of ordinary spherical coordinate evolutions of a
strongly perturbed Minkowski spacetime. In
Sec.~\ref{sec:convergence_tests}, we demonstrate in the contexts of a
single, nonspinning black hole and a double black hole head-on
collision that the finite difference truncation error converges to
zero with increasing grid resolution at the expected rate, and that
increasing finite difference order with fixed grid resolution results
in near-exponential convergence of the error.

These tests also act to showcase the efficiency of SENR/NRPy+ against
other open-source numerical relativity codes in the context these
physical scenarios, as all tests were performed on desktop-scale
computers, using at most only about $1.5$~GB of RAM.

\subsection{Code comparisons}
\label{sec:code_comparisons}

Here we directly compare SENR/NRPy+ results with two other
established BSSN evolution codes. These tests verify that all of the
evolution equations and diagnostics are implemented correctly, and
that the simulations do indeed contain black hole horizons with the
expected properties.

\subsubsection{Robust stability test: Roundoff-level agreement between SENR/NRPy+ and BMCCM}

This section compares SENR/NRPy+ to the spherical-polar, spatially fourth-order
finite differenced BSSN code of Baumgarte \emph{et al.}~\cite{Baumgarte:2013}
(hereafter BMCCM). BMCCM includes many features beyond the scope of
this paper~\cite{Baumgarte:2015a, Baumgarte:2015b, Baumgarte:2016,
  Gundlach:2016, Miller:2017}; here we focus on initial data that
represent a version of the robust stability 
test bed~\cite{Szilagyi:2000, Szilagyi:2002,Alcubierre:2004,
  ApplesRobustStabilityTest}, which involves a strong random
perturbation about flat spacetime. At each grid point, the random
number generator \texttt{drand48}~\cite{Roberts:1982} is seeded with a
unique (but constant) integer tied to the grid index. Then, each of
the grid functions are populated, in turn, with Minkowski initial data
($e^{\phi} = 1$, $\varepsilon_{i j} = 0$, $\bar{A}_{i j} = 0$, 
$K = 0$, $\alpha = 1$, $\beta^{i} = 0$, and $B^{i} = 0$) plus a random value picked from
the uniform distribution $[-0.02, 0.02]$. This produces repeatable
initial data with no spatial correlation, and tests every aspect of
the evolution and diagnostic algorithms.

Standard second-order Runge-Kutta (RK2) integration is unconditionally
unstable in the presence of coordinate
singularities. To circumvent this problem, BMCCM
evolves the BSSN fields using the second-order partially implicit
Runge-Kutta (PIRK2) integrator~\cite{Cordero-Carrion:2014}
(see~\cite{Cordero-Carrion:2012} for details of the method and its
higher-order generalizations). The PIRK2 method treats all of the
regular terms in the evolution equations explicitly, as in RK2, but
evolves the singular terms by an implicit step that depends on the
updated values of the regular parts. This technique does not require
any analytical or numerical inversion, and so its cost is comparable
to that of fully explicit schemes. 
Although by default SENR implements RK4, which {\it is} stable in the
presence of coordinate
singularities~\cite{Cordero-Carrion:2012}, we
also implement PIRK2 in SENR to directly compare results with the
established BMCCM code.

We compare between SENR/NRPy+ and BMCCM the $L_{2}$ norm of the
Hamiltonian violation over the entire grid, distilling an entire
grid's worth of data down to a single number at each
time. Double precision floating point arithmetic maintains
approximately 16 digits of significance in any mathematical operation,
limiting the extent to which it can be said that two algorithms are in
numerical agreement. Over the course of a simulation, errors at
the 16th significant digit will gradually rise---an unavoidable
phenomenon when using finite-precision arithmetic known as roundoff
error. If two codes are shown to produce results that agree to within
roundoff error, they are functionally identical, so that if one code
has been proven robust, then the other code possesses identical
robustness.

\begin{figure}[!t]
\centering
\includegraphics[width=\columnwidth]{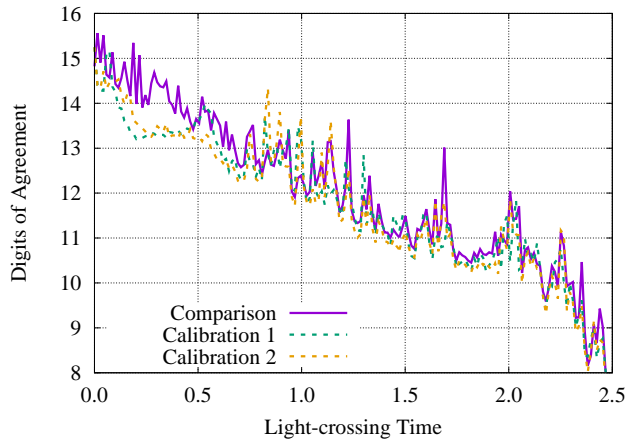}
\caption{Digits of agreement between SENR/NRPy+ and BMCCM evolving
  random initial data, measuring the $L_{2}$ norm of the Hamiltonian
  constraint over the entire grid. The initial data are a random
  perturbation about flat spacetime of maximum magnitude $0.02$. The
  calibration runs compare BMCCM with itself randomly perturbed,
  initially, in the least significant digit. Both codes terminate due
  to NaNs at exactly the same iteration.}
\label{fig:roundoff_comparison}
\end{figure}

The comparison is calibrated by first running BMCCM with the random
initial data. Then, it is run again with identical initial data,
except that the least significant digit at every point is reset to a random
number. This tiny initial difference grows over time due to roundoff
error. Given the strong, discontinuous nature of the random perturbations away
from flat space, infinities and NaNs eventually develop, causing the
simulations to terminate. For good measure, we perform the calibration
a second time, starting with an identical perturbation of
Minkowski, but re-randomizing the least significant digit for
each function at every grid point. The results diverge from the base
case in a similar fashion, shown as dashed lines in
Fig~\ref{fig:roundoff_comparison}.

Finally, we run the initial data of the BMCCM base case through
SENR/NRPy+ on an identical grid. The resulting differences are
illustrated as a solid line in
Fig~\ref{fig:roundoff_comparison}. SENR/NRPy+ maintains significance
at least as well as the BMCCM calibrations, indicating the two codes
agree to within roundoff error. Thus the results of the
SENR/NRPy+ and BMCCM codes are numerically indistinguishable.

\subsubsection{SENR/NRPy+ and the Einstein Toolkit: Comparison of
  nonspinning puncture black hole evolutions}
\label{sec:SENR_ET}

In this section, results from SENR/NRPy+ are compared with the
ETK in the context of nonspinning black hole
evolutions, tracking the apparent horizon radii.
The initial data represent a single wormhole slice of the
Schwarzschild black hole spacetime in isotropic coordinates. These
data are conformally flat ($\varepsilon_{i j} = 0$), maximally sliced
($K = 0$), and exist at a moment of time symmetry 
($\bar{A}_{i j} = 0$). The conformal factor is the solution to the
flat space Laplace equation when the asymptotic flatness condition 
$\lim_{r \to \infty} e^{\phi} = 1$ is imposed at infinity
\beq
  \label{eq:wormhole_conformal_factor}
  e^{\phi} = 1 + \frac{M}{2 r} \; ,
\eeq
where $r$ is the isotropic radial coordinate distance to the puncture
located at the origin. The constant of integration is chosen such that
the total ADM energy~\eqref{eq:ADM_Mass} of the slice is
$M_{\text{ADM}} = M$. As initial gauge conditions, we use the popular
``pre-collapsed'' lapse~\cite{Alcubierre:2003b}
\beq
  \label{eq:pre_collapsed_lapse}
  \alpha = e^{-2 \phi}
\eeq
and vanishing shift $\beta^{i} = 0$ and $B^{i} = 0$. The shift
evolution damping parameter $\eta$~\eqref{eq:Gamma-driver_B} influences
the coordinate radius of the black hole apparent horizon.
We found that our choice of $\eta = 0.25/M$ results in a horizon
that quickly settles down to a static state.

In the ETK simulation, we make use of the open-source
Cactus/Carpet~\cite{Cactus,Carpet} AMR
infrastructure to place a single wormhole black hole at the
origin, surrounded by 8 levels of mesh refinement. The AMR grids
are Cartesian and adopt a Cartesian basis. Each refinement level
contains the same number of points (excluding ghost and AMR buffer
zones), and the grid spacing doubles each time a refinement boundary is
crossed (starting from the origin and moving outward). The grid
outer boundary is set to $r_{\text{max}} / M = 128$. There are 32 uniformly spaced grid points
across each refinement level in each coordinate direction, giving 16
points across the horizon when at its smallest (on the initial slice). On the finest
refinement level, the grid spacing along each of the coordinate
directions is $\Delta r_{\text{min}} / M = 0.03125$. RK4 time
evolution (the method of lines) and sixth-order finite difference stencils are adopted, with
upwinding on the shift-advection terms. The BSSN equation C code in
the McLachlan BSSN thorn~\cite{Loffler:2012} is
automatically generated using the Mathematica-based Kranc code~\cite{Husa:2006}.
The black hole apparent horizon radius is measured from the evolved fields using
the AHFinderDirect thorn~\cite{Thornburg:2004a}.

In SENR/NRPy+, we adopt the same initial data, but place it on a
single spherical grid with coordinate redistribution function
\beq
  \label{eq:sinh_radial}
  f(\xOne) = r_{\text{max}} \frac{\sinh\left(\xOne / w\right)}{\sinh\left(1 / w\right)} \; ,
\eeq
with all tensorial variables expressed in the spherical basis.
The grid parameters are tuned to match both the outer
boundary of the ETK simulation, as well as the minimum
grid spacing at the black hole. The ETK AMR grid has $32 \times 8 = 256$ 
($192$ nonoverlapping) points along each Cartesian coordinate direction. 
Along the diagonal, the resolution is effectively reduced to
$256 / \sqrt{3} \approx 148$ points. For the SENR grid, we allocate 
$N_{\xOne} = 148$ and $N_{\xTwo} = N_{\xThree} = 2$ 
(the minimum number of angular grid points required
by our boundary condition module), let $w = 0.173435$, and set the outer
boundary $r_{\text{max}} / M = 128$. This results in 
$\Delta r_{\text{min}} / M \approx 0.03123$, agreeing well with the ETK grid.
The black hole apparent horizon radius is measured from the evolved fields
by hunting for the root of the null expansion (see Sec.~\ref{sec:NRPy_diagnostics} for details).
We adopt $N_{\text{FD}} = 6$ and RK4 time integration in SENR/NRPy+ to match ETK.

Despite tuning the grid resolutions and basic numerical evolution
strategies to be consistent across codes, the chosen shift condition
in both codes is not covariant [note the partial derivatives
appearing in Eqs.~\eqref{eq:Gamma-driver} and~\eqref{eq:partial0},
and see discussion in Brown~\cite{Brown:2009} for how to make
this shift condition covariant]. Thus we should not in general expect
results that agree between the codes, as they do not adopt
the same coordinate system. For spherically symmetric spacetimes,
however, the partial derivatives $\partial_i$ in
Eqs.~\eqref{eq:Gamma-driver} and~\eqref{eq:partial0} can be replaced
with the covariant derivative $\hat{D}_i$ in both spherical and
Cartesian coordinates, showing that, in this special case, the gauge
conditions are geometrically identical.

\begin{figure}[!t]
\centering
\includegraphics[width=\columnwidth]{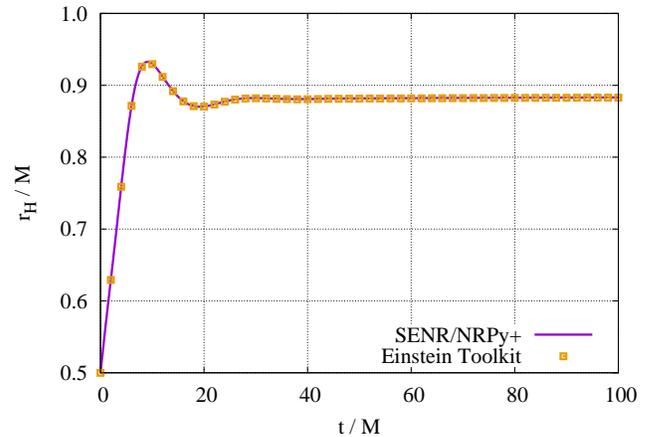}
\caption{Single black hole wormhole initial data evolved in the standard gauge
  with sixth-order finite differencing: comparison between SENR/NRPy+ and the
  ETK. Evolution of the apparent horizon coordinate radius as measured in
  SENR/NRPy+ by finding the root of the null
  expansion~\eqref{eq:null_expansion}, and in the ETK using the
  AHFinderDirect thorn~\cite{Thornburg:2004a}.}
\label{fig:ET_SENR_comparison}
\end{figure}

In Fig.~\ref{fig:ET_SENR_comparison}, we monitor the coordinate
radius of a puncture black hole as an indicator of the spacetime field
and shift dynamics (obviously, the shift condition directly influences
the coordinate radius of a puncture black hole). Notice that the
apparent horizon radii measured by SENR/NRPy+ 
and ETK {\it match extremely well over time}, starting from the
expected initial coordinate radius of the wormhole throat
$r_{\text{H}} / M = 0.5$ and equilibrating to a trumpet coordinate
radius of $r_{\text{H}} / M \approx 0.883$.

We find that the constraint violation in SENR/NRPy+ versus radius is typically
below the level observed in the ETK simulation by about 2 orders of
magnitude, and at worst the two share the same level of violation.

Even though results between the two codes in the
strong-field region agree extremely well (Fig.~\ref{fig:ET_SENR_comparison}), and the magnitude of
constraint violations is significantly smaller in SENR/NRPy+, the ETK
simulation requires approximately 10~GB of RAM, whereas the SENR/NRPy+
simulation needs only 28~MB---about 0.28\% of the ETK simulation. Of
course this is due to the ability of SENR/NRPy+ to take advantage of the
spherical symmetry in the spherical basis, while the ETK simulation
models the black hole on nested Cartesian AMR grids. Note that in
spite of the spherical symmetry, this is still a full 3+1 simulation
for SENR, just with very few points sampling the angular directions.

Having demonstrated excellent agreement with the BMCCM and ETK codes,
we next turn our attention to code validation tests, in which numerical
errors in SENR/NRPy+ are demonstrated to converge to zero as expected.

\subsection{Convergence tests}
\label{sec:convergence_tests}

Section~\ref{sec:grid_convergence} shows that finite difference truncation errors converge
to zero with increasing grid resolution at the expected rates in
SENR/NRPy+. We show in Sec.~\ref{sec:trumpet_convergence} that
the truncation error converges to
zero nearly exponentially with linear increase in the finite
difference order, keeping the numerical grids held fixed. We explore the
convergence behaviors of physical quantities extracted from the
evolved fields. Then, in Sec.~\ref{sec:head-on} we evolve the dynamical head-on
collision from rest of two nonspinning black holes, and confirm that
the ringdown of the merged black hole gravitational waveform matches
the analytical prediction both in frequency and amplitude.

\subsubsection{Convergence of puncture black hole evolutions}
\label{sec:grid_convergence}

The BSSN equations are solved on the uniformly sampled
$x^{(i)}$ grid, which directly maps to
the solution on the nonuniformly sampled, Cartesian $y^{(i)}$
grid. In particular, we use the linear coordinate redistribution function
\beq
  \label{eq:linear_redistribution}
  f(\xOne) = r_{\text{max}} \, \xOne \; ,
\eeq
and similarly for $\xTwo$ or $\xThree$, depending on the choice of coordinates.
Numerical errors in solving these equations stem largely from
the finite difference representation of spatial derivatives (i.e.,
truncation errors of these derivatives are typically dominant).
Finite difference operators effectively fit a polynomial to a function
sampled at a fixed number of neighboring points, so that the
derivative of the polynomial acts as an approximation to the exact
derivative. Truncation error---i.e., the error caused by approximating
functions with finite polynomials of degree $D$---drops as the sample
rate to some power $N_{\text{FD}}$ that is related to $D$. In our
finite difference schemes, $N_{\text{FD}} = D$.
 
In this section, we confirm that in fact the error drops in proportion
to our underlying uniform grid spacing $\mathcal{E}_{\text{FD}} \sim \mathcal{O}\left(\Delta x^{N_{\text{FD}}}\right)$, in the
context of a single, nonspinning puncture black hole evolved using the
BSSN formalism, in which $N_{\text{FD}} \in \{2, 4, 6\}$. We monitor the Hamiltonian
constraint violation. With uniform
resolution, it becomes very expensive to simultaneously resolve the
apparent horizon and push the outer boundary far from the
puncture. This limits the total run time before error from the
outer boundary contaminates the horizon. 

As an alternative approach, we perform convergence tests in which the
grid resolution is held fixed and the finite difference order is
allowed to vary. In this way, we increase the degree of the finite difference polynomial
that is fit to each function at each point. Therefore, we
expect the convergence rate to be approximately exponential,
provided roundoff error is sufficiently small
and the underlying functions are smooth (see, e.g.,~\cite{Boyd:2001}
for additional discussion of ``exponential'' convergence). 

These tests adopt the same isotropic wormhole initial data and initial
gauge conditions as in Sec.~\ref{sec:SENR_ET}. We set the outer
boundary to $r_{\text{max}} / M = 10$. The fastest waves on the grid (related to
the 1+log lapse condition; see,
e.g.,~\cite{vanMeter:2006, Etienne:2014})
propagate, with speed $\sqrt{2 \alpha}$, inward from the outer boundary and
outward from the puncture. These simulations end at $t / M \approx 5$. 

\begin{figure}[!t]
\centering
\includegraphics[width=\columnwidth]{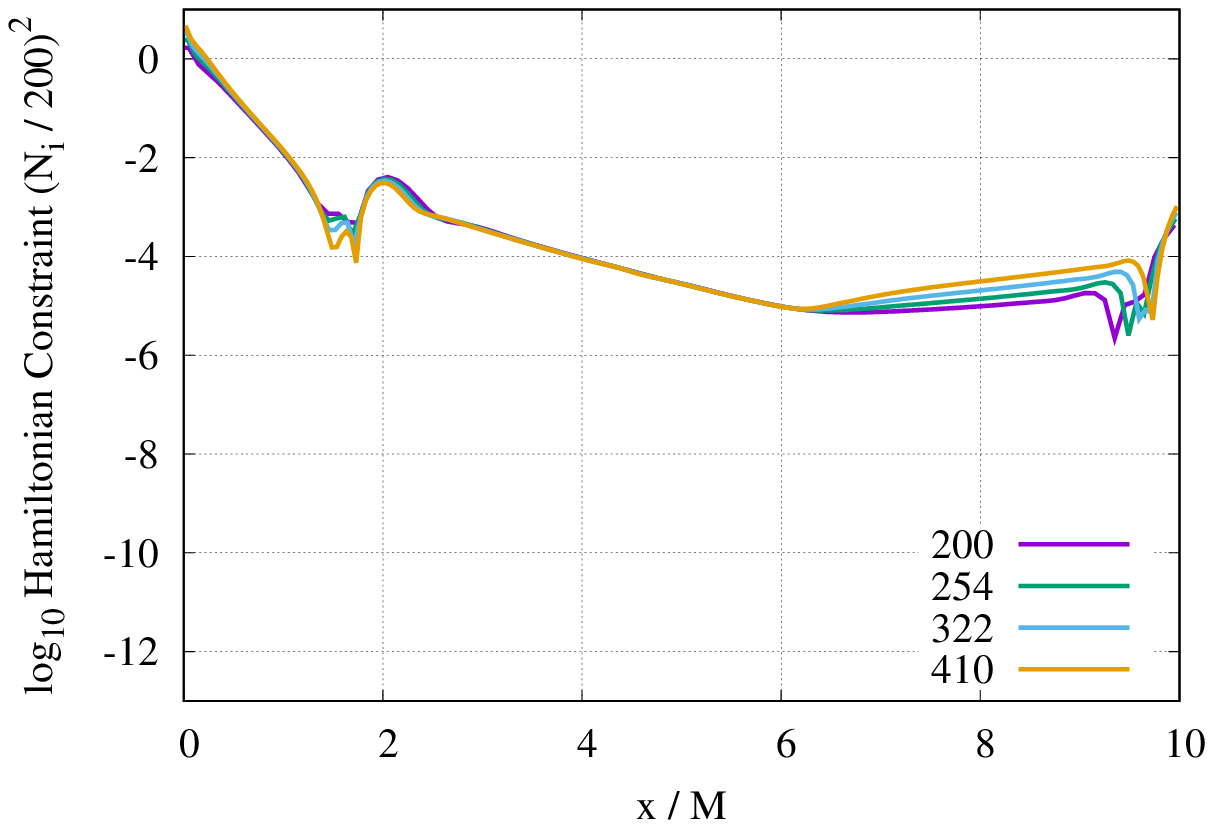}
\includegraphics[width=\columnwidth]{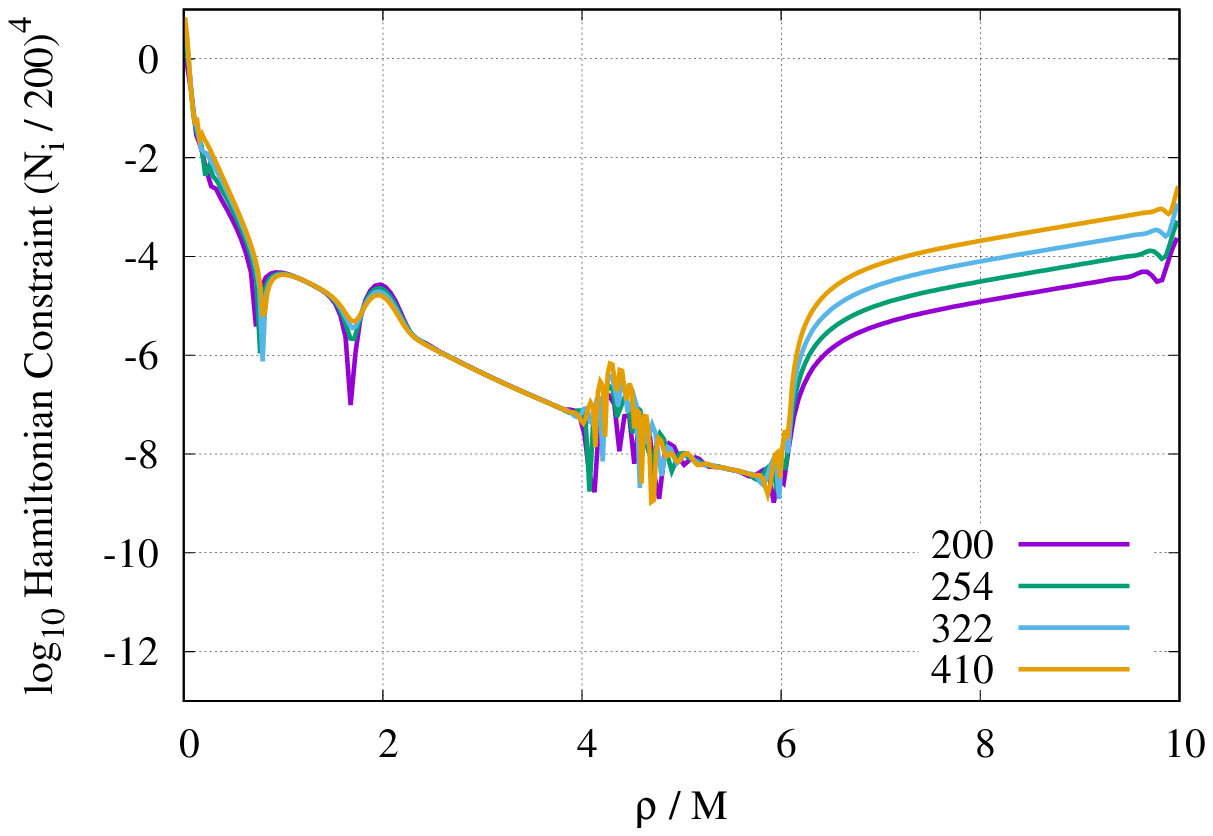}
\includegraphics[width=\columnwidth]{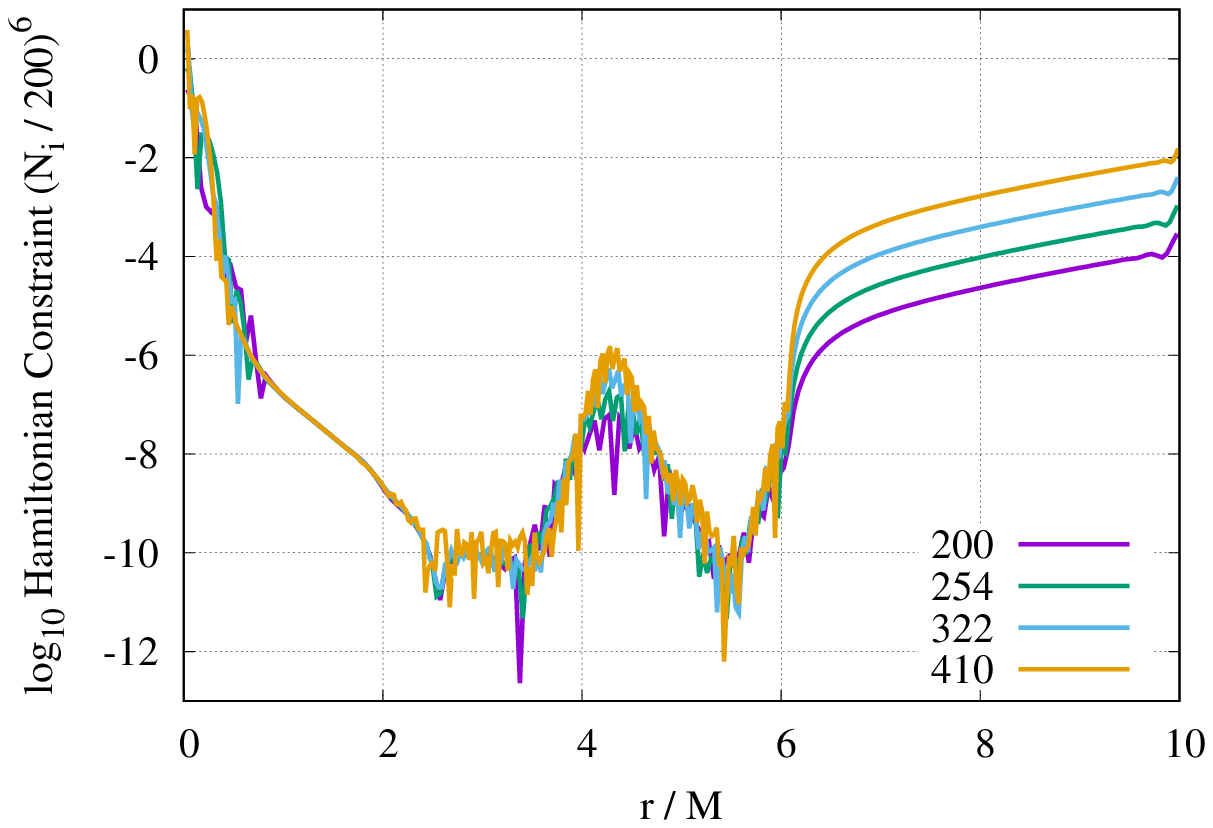}
\caption{Single black hole wormhole initial data evolved in the standard gauge: convergence of
  truncation errors to zero of Hamiltonian constraint violation in linearly distributed
  Cartesian ({\bf top panel}), cylindrical ({\bf middle panel}), and
  spherical coordinates ({\bf bottom panel}). The legends indicate the number of
  grid points $N_{i}$ in each (nonangular) direction. The data shown here are scaled 
  by the factor $(N_{i} / 200)^{N_{\text{FD}}}$.
  Data are measured along a radial line at 
  $t / M \approx 5$. Cartesian, cylindrical, and spherical
  coordinate evolutions are performed with $N_{\text{FD}} = 2$, 4, and
  6, respectively. We adopt the linear redistribution function parameter
  $r_{\text{max}} / M = 10$ [Eq.~\eqref{eq:linear_redistribution}].}
\label{fig:grid_convergence}
\end{figure}

In Fig.~\ref{fig:grid_convergence}, we show that the
finite difference truncation errors converge at the expected rates in
all three coordinate system classes by rescaling the higher resolution
data by a constant factor $(N_{i} / 200)^{N_{\text{FD}}}$, where
$N_{i} \in \{200, 254, 322, 410\}$ represents the number of grid points 
(in the nonangular coordinate directions) in each run. 
We let $N_{\text{FD}} = 2$ in the Cartesian case,
$N_{\text{FD}} = 4$ in the cylindrical case, and $N_{\text{FD}} = 6$
in the spherical case. The lack of convergence at $r / M \gtrsim 6$
is due to the approximate outer boundary conditions; in practice
we push outer boundaries out of contact from the physical system of
interest via a simple logarithmic radial rescaling of the underlying
coordinate system.

\begin{figure}[!t]
\centering
\includegraphics[width=\columnwidth]{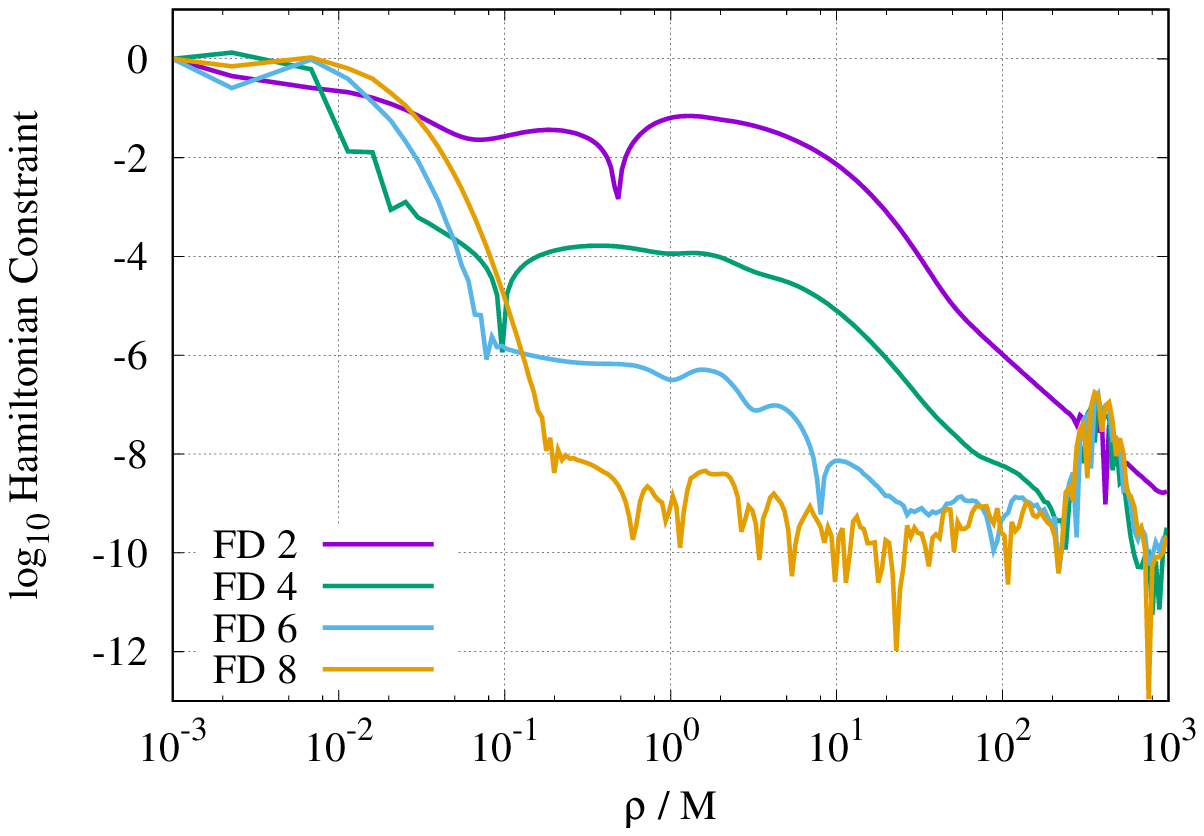}
\includegraphics[width=\columnwidth]{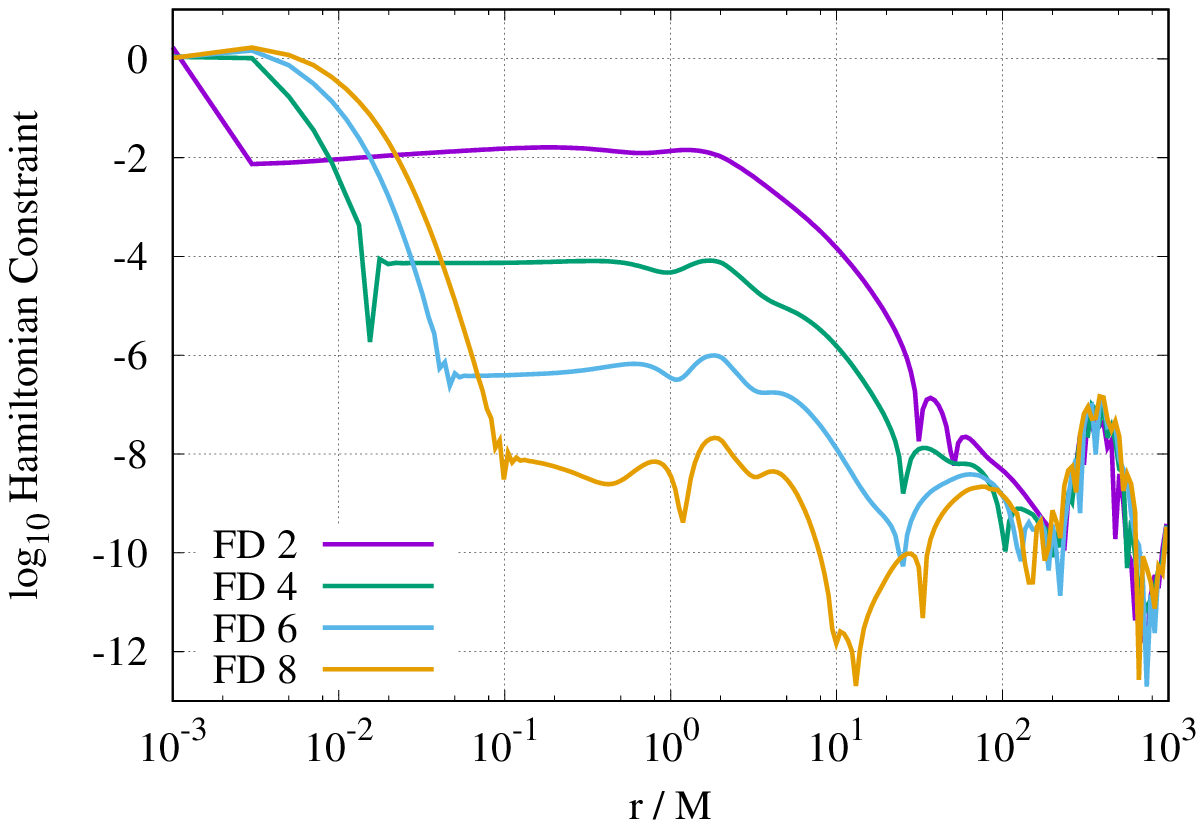}
\caption{Single black hole wormhole initial data evolved in the standard gauge: convergence of
  truncation errors to zero of Hamiltonian constraint violation in
  sinh-cylindrical ({\bf top panel}) and sinh-spherical ({\bf bottom
    panel}) coordinates. Numerical grids are held fixed at moderate
  resolution, and only finite difference order is increased.
  Hamiltonian constraint violation is measured along a radial line at
  $t / M \approx 300$. As the nonconvergent gauge wave pulse propagates
  outward from the puncture, exponential convergence of the
  Hamiltonian is restored in its wake. For these simulations, we take
  ${\eta = 2 / M}$. We adopt redistribution parameters 
  $r_{\text{max}} / M = 1000$  and ${w = 0.0916845}$ [Eq.~\eqref{eq:sinh_radial}]. 
  The spherical-like grid uses $N_{\xOne} = 200$ and $N_{\xTwo} = N_{\xThree} = 2$ points,
  and the cylindrical-like grid uses $N_{\xOne} = 200$, $N_{\xTwo} = 2$, and $N_{\xThree} = 400$.}
\label{fig:FD_convergence}
\end{figure}

Remarkably, in Fig.~\ref{fig:FD_convergence}, we
find that while the Hamiltonian constraint violation is 
{\it anticonvergent} inside the horizon ($r / M \lesssim 0.5$), we
maintain approximately exponential convergence in a sizable region {\it outside} the
horizon, hitting roundoff error at eighth order.
We have observed nonconvergent behavior propagating outside
the horizon only in cases when the total number of points inside the
horizon are set to be so small that finite difference stencils outside
the horizon touch grid points immediately surrounding the puncture.

More alarming than the puncture itself, the sharp lapse wave (a
``gauge shock''~\cite{Alcubierre:2003a, Alcubierre:2005}) that propagates
outward from the horizon in moving puncture simulations is another source of
nonconvergence.  Remarkably, exponential-like convergence is restored
in the wake of this gauge wave pulse; in Fig~\ref{fig:FD_convergence},
the pulse has reached $r / M \approx 425$ at the time of
measurement. Restoration of convergence after the gauge pulse is
very difficult to achieve on Cartesian AMR grids, as sharp
outgoing waves are partially reflected off of refinement
boundaries. (See, e.g.,~\cite{Etienne:2014} for discussion of how this
problem might be mitigated.) 

Based on these results, we anticipate much cleaner exponential-like
convergence in black hole evolutions for which this gauge pulse does
not exist. Next we explore just such a case: evolutions of trumpet
black hole initial data.

\subsubsection{Convergence of evolved static trumpet initial data}
\label{sec:trumpet_convergence}

The trumpet solution represents a time-independent slicing of the
Schwarzschild spacetime~\cite{Dennison:2014}. In this section, we adopt the trumpet
solution to show that the numerical evolution
outside the horizon is completely dominated by truncation error.

The trumpet data are conformally flat ($\varepsilon_{i j} = 0$), and
describe a single black hole with mass $M$. With the choice $f_{1} =
R_{0} = M$ in~\cite{Dennison:2014}, the trumpet conformal factor is
\beq
  \label{eq:trumpet_conformal_factor}
  e^{\phi} = \sqrt{1 + \frac{M}{r}}
\eeq
and the nonvanishing extrinsic curvature terms are
\begin{subequations}
  \begin{align}
    K &= \frac{M}{(M + r)^{2}} \; , \\
    \bar{A}_{r r} &= - \frac{4 M}{3 (M + r)^{2}} \; , \\
    \bar{A}_{\theta \theta} = \frac{\bar{A}_{\varphi \varphi}}{\sin^{2}(\theta)} &= \frac{2 M r^{2}}{3 (M + r)^{2}} \; .
  \end{align}
\end{subequations}

Using an alternative to the standard 1+log condition given by
Eq.~\eqref{eq:1_plus_log}, the lapse is evolved according to a condition
consistent with staticity 
\beq
  \partial_{0} \alpha = - \alpha (1 - \alpha) K \; .
\eeq

For the shift vector evolution equation, we desire only that the
right-hand sides vanish analytically (although numerical error is
expected to result in specious evolution). To this end, we adopt the
nonadvecting Gamma-driver condition 
\begin{subequations}
\begin{align}
  \partial_{t} \beta^{i} &= B^{i} \; , \\
  \partial_{t} B^{i} &= \frac{3}{4} \partial_{t} \bar{\Lambda}^{i} - \eta B^{i} \; .
\end{align}
\end{subequations}
The initial lapse and shift take on the forms
\begin{subequations}
  \label{eq:trumpet_lapse_shift}
  \begin{align}
    \alpha &= \frac{r}{M + r} \; , \\
    \beta^{r} &= \frac{M r}{(M + r)^{2}} \; .
  \end{align}
\end{subequations}
We use damping parameter $\eta = 0.25/M$, although the
results are not very sensitive to its particular
value because the evolution begins and remains in a quasistatic state.

Analytically, the trumpet solution with these gauge conditions is
static, but numerical errors result in unwanted evolution of the
fields away from the initial data. We perform numerical evolutions of
these data on fixed numerical grids, subject to these gauge evolution equations, to confirm that
evolution away from the initial data disappears nearly exponentially with increased finite
difference order.

We choose a fixed, spherical-like
coordinate grid with 
$N_{\xOne} = 128$ and $N_{\xTwo} = N_{\xThree} = 2$, where the
radial points are distributed according to Eq.~\eqref{eq:sinh_radial} with
$w = 0.0747$ and $r_{\text{max}} / M = 1000$ (the location of the
outer boundary). We perform numerical evolutions on these fixed grids
at finite difference orders $N_{\text{FD}} \in \{2, 4, 6, 8\}$.

\begin{figure}[!t]
  \centering
  \includegraphics[width=\columnwidth]{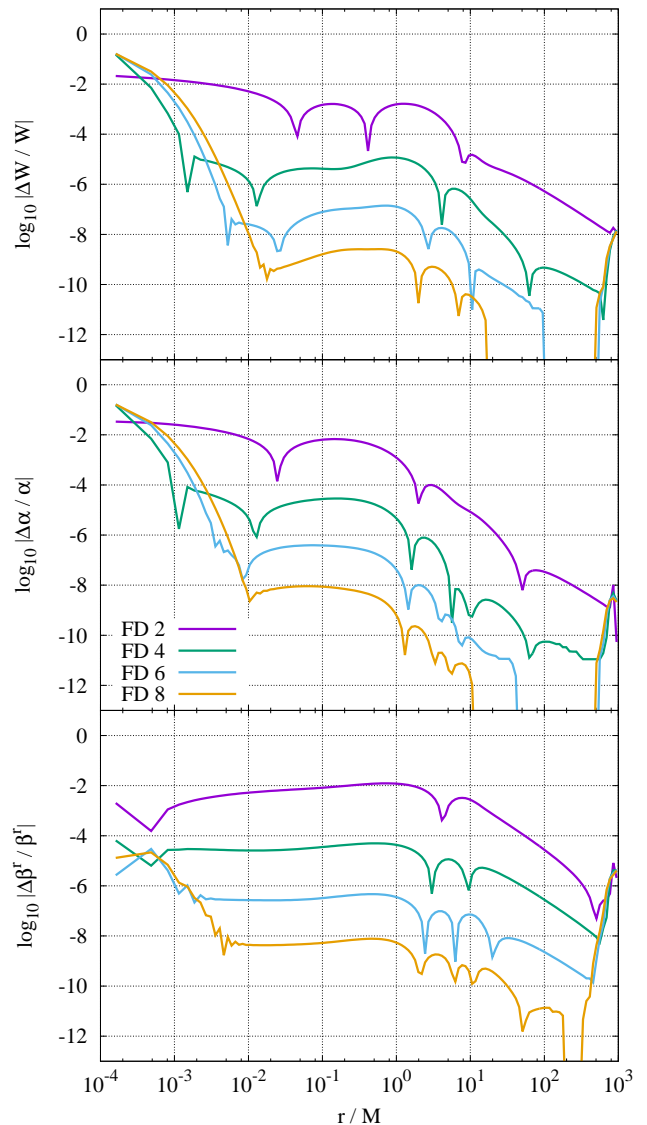}
  \caption{Trumpet black hole initial data evolved for ${t / M \approx 100}$ in the 
    static trumpet gauge: relative
    difference between evolved and initial conformal factor $W$ ({\bf top panel}),
    lapse $\alpha$ ({\bf middle panel}), and radial
    component of the shift $\beta^{r}$ ({\bf bottom panel}), versus
    distance from the origin. The trumpet black hole is centered on
    the origin.}
  \label{fig:StaticTrumpet_FD_convergence}
\end{figure}

After $t / M \approx 100$, the freely evolved
conformal factor $W$, and the gauge functions $\alpha$ and
$\beta^{r}$, are compared with their initial values. The relative
differences are shown in Fig.~\ref{fig:StaticTrumpet_FD_convergence}
for varying $N_{\text{FD}}$. As in the puncture evolution of the
previous section, we observe nonconvergent numerical errors inside
the horizon ($r/M\lesssim 0.5$), which leads to additional spurious
dynamics in the black hole with increased finite difference
order. Unlike in the 1+log evolutions, however, no sharp lapse
wave exists in the trumpet solution, so that exponential convergence is
maintained over the entire numerical grid outside the horizon and
inside the region influenced by the approximate outer boundary
conditions. In fact, at high finite difference order, all plotted quantities
drop to roundoff-level agreement somewhere in the region between
$r / M = 10$ and $100$.

\subsubsection{Gravitational waves from a head-on collision}
\label{sec:head-on}

In Sec.~\ref{sec:grid_convergence}, we demonstrated that nearly exponential
convergence outside a puncture black hole horizon (and inside the
region causally influenced by the outer boundary) is restored in the
wake of a sharp gauge wave. In~\cite{Zlochower:2012, Etienne:2014}, it
is posited that this sharp wave causes nonconvergent errors
in moving puncture black hole binary simulations on AMR grids, due to
reflections off refinement boundaries. These nonconvergent errors
have a direct impact on the convergence of the gravitational waves
in these simulations.

In this section, we explore the convergence of gravitational wave
signals from a head-on Brill-Lindquist black hole collision on
spherical-like coordinate grids, keeping the numerical grids fixed at
a moderate resolution and varying only finite difference derivative
order, choosing $N_{\text{FD}} \in \{2, 4, 6, 8, 10\}$. 

Brill-Lindquist initial data~\cite{Brill:1963} represent nonspinning
black holes starting from rest. The initial data are constructed from
a superposition of isotropic wormhole slices of the Schwarzschild
spacetime. Though the formulation holds for an arbitrary number of
black holes, here we use only two.

The wormhole slice of the Schwarzschild spacetime with isotropic
radial coordinate $r$ is conformally flat ($\varepsilon_{i j} = 0$) on
the initial Cauchy surface. This hypersurface is maximal ($K = 0$) and
exists at a moment of time symmetry ($\bar{A}_{i j} = 0$). The conformal
factor is the solution to the flat-space Laplace equation, which
allows for a direct superposition of single wormhole conformal
factors given in Eq.~\eqref{eq:wormhole_conformal_factor}
\beq
  e^{\phi} = 1 + \frac{M_{+}}{2 r_{+}} + \frac{M_{-}}{2 r_{-}} \; ,
\eeq
where
\beq
  r_{\pm} = \sqrt{r^{2} + b^{2} \mp 2 b r \cos(\theta)}
\eeq
is the isotropic radial coordinate distance from the coordinate origin
to the puncture with mass parameter $M_{\pm}$ located along the
spherical-polar axis $\pm b / M$ above ($+$) or below ($-$) the
origin. This configuration is axisymmetric with respect to the polar
axis, though, as with all of the other cases, we perform our simulation
in full 3+1 dimensions.

For the runs presented here, we focus on an equal-mass case in which
$M_{\pm} = M / 2$, so that the ADM mass integral~\eqref{eq:ADM_Mass}
gives $M_{\text{ADM}} = M$. We confirm that the ADM
momentum~\eqref{eq:ADM_Momentum} and angular
momentum~\eqref{eq:ADM_Angular_Momentum} integrals vanish, as
expected. We also chose $b / M = 0.5$.

We use Eq.~\eqref{eq:pre_collapsed_lapse} for the initial lapse, and $\beta^{i} = 0$ 
and $B^{i} = 0$ for the initial shift. The shift evolution damping parameter 
is set to $\eta = 2 / M$, which results in an equilibrium remnant horizon radius
of $r_{\text{H}} / M \approx 1.4$.
We use a spherical-like grid with redistribution function
equation~\eqref{eq:sinh_radial}, so that $\xOne$ corresponds to a radial
coordinate. We choose $r_{\text{max}} / M = 1000$ and 
$w = 0.125$ with $N_{\xOne} = 400$, $N_{\xTwo} = 64$, and $N_{\xThree} = 2$ (where
$N_{\xThree}$ is the axis of symmetry for the collision). Evolving
wormhole initial data with the ordinary 1+log lapse
condition Eq.~\eqref{eq:1_plus_log} results in a sharp, nonconvergent gauge
pulse that propagates outward from the puncture. We find, however,
that excellent convergence is restored in the wake of the pulse as it moves
towards the outer boundary.

\begin{figure}[!t]
\centering
\includegraphics[width=\columnwidth]{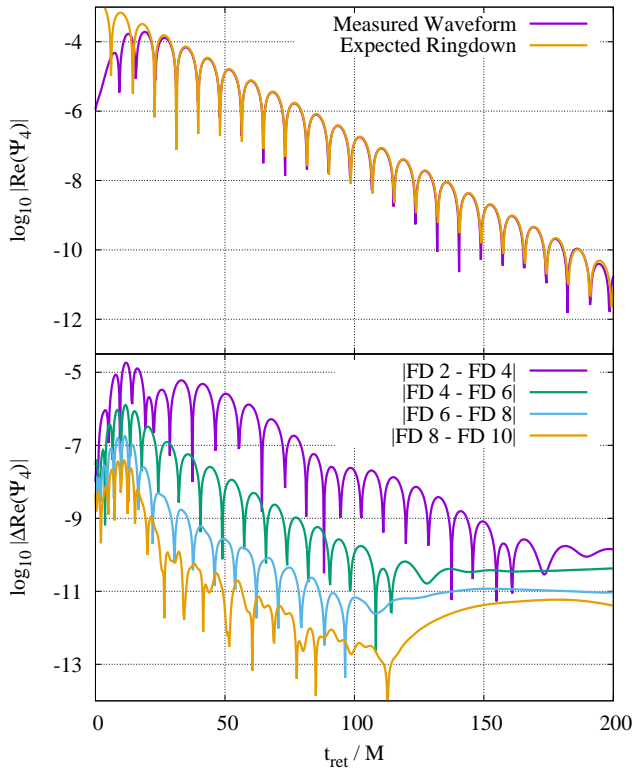}
\caption{Analysis of gravitational waves from the head-on collision of
  two puncture black holes resulting from Brill-Lindquist initial data
  evolved with the standard gauge conditions. The common axis shows the retarded time
  $t_{\text{ret}} \equiv t - r_{\text{ext}}$, where the gravitational wave
  extraction radius is $r_{\text{ext}} / M \approx 44.8$. 
  ({\bf Top panel}) Dominant $(\ell = 2, m = 0)$ mode of $\Psi_4$,
  $|\Re(\Psi_{4})|_{\ell=2,m=0}$, plotted atop
  $\left|\Psi^{\rm rd}_{4}\right|$, the quasinormal mode ringdown
  amplitude and frequency expected from black hole perturbation theory
  [Eq.~\eqref{eq:expectedringdownPsi4}]. Numerical results are plotted at finite difference
  order $N_{\text{FD}} = 10$.
  ({\bf Bottom panel}) Convergence of absolute differences in
  $|\Re(\Psi_{4})|_{\ell=2,m=0}$ at adjacent finite differencing
  orders, keeping the spherical grid fixed at moderate resolution.} 
\label{fig:gw_FD_convergence}
\end{figure}

The black holes are both nonspinning and are released from rest, so
they remain on the polar axis during infall and collide
head-on. They merge to form a single, strongly perturbed black hole,
which quickly rings down to a stationary state as the time-changing
quadrupole in the horizon is radiated away in the form of
gravitational waves. The result is that the waveform after merger
behaves as an exponentially damped harmonic oscillator.
In spin-weight-two spheroidal harmonics, the fundamental gravitational
wave mode in the $\ell = 2$ harmonic has complex frequency 
$\omega M \approx 0.3737 - 0.0890 i$~\cite{Berti:2009}. With only a
constant amplitude $A_{\text{f}}$ and phase offset $\phi_{\text{f}}$ acting as fitting
parameters, the magnitude of the expected ringdown signal
\beq
\Re\left(\Psi^{\rm rd}_{4}\right) = A_{\text{f}} \exp(-0.0890\, t) \cos(0.3737\, t + \phi_{\text{f}})
\label{eq:expectedringdownPsi4}
\eeq
is plotted in the upper panel of Fig.~\ref{fig:gw_FD_convergence},
atop the amplitude of the real part of the dominant $\ell=2$ mode of
$\Psi_{4}$ measured in our simulation. Excellent agreement is observed
between the expected ringdown signal and the results from our
simulation {\it over more than six decades} in amplitude. Further, the
symmetry of the head-on collision is expected to result in
gravitational waves that are in a pure $+$ polarization state, which
we confirm by measuring the imaginary part of $\Psi_{4}$ to be zero to
roundoff error [see Eq.~\eqref{eq:Weyl_scalar}].

The bottom panel of Fig.~\ref{fig:gw_FD_convergence} demonstrates that
differences in waveforms at adjacent finite difference orders
(keeping the spherical-like coordinate grids fixed at moderate
resolution) converge nearly exponentially with increased
finite difference order. Notice that after the peak gravitational
wave signal has passed, $|\text{FD 6} - \text{FD 8}|$ and
$|\text{FD 8} - \text{FD 10}|$ are at times influenced by roundoff
error, as evidenced by their suddenly stochastic behavior. We
confirmed this feature by repeating the simulations with
{\tt long double} (80-bit) floating-point precision.

The peak amplitude occurs at retarded time of approximately
$t_{\rm ret} / M \approx 18$, when the differences between waveforms
at adjacent finite difference orders are near their peaks. At this
time, this mode of $\Psi_{4}$ gains about an order of magnitude more
precision with each increment of finite difference order. 
The particular rate of exponential convergence depends on the grid spacing.

\begin{figure}[!t]
\centering
\includegraphics[width=\columnwidth]{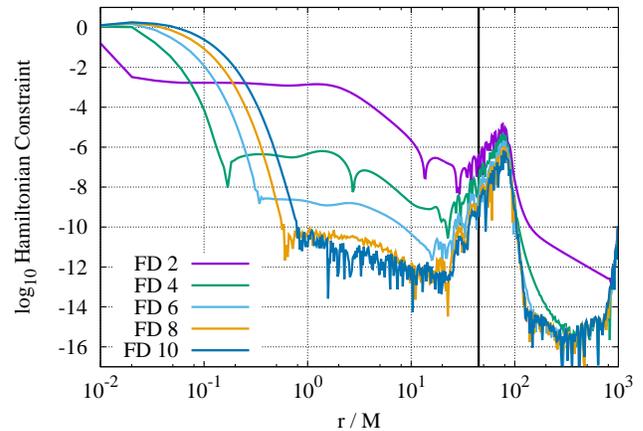}
\caption{Brill-Lindquist head-on collision of two puncture black holes: Hamiltonian
  constraint violation at ${t / M \approx 62.9}$, such that the peak gravitational
  wave amplitude passes through $r_{\text{ext}} / M \approx 44.8$, the radius at which
  gravitational waves are extracted in
  Fig.~\ref{fig:gw_FD_convergence}. Spherical grids are fixed at
  moderate resolution; only finite differencing order is varied. The black
  vertical line denotes the approximate location of the peak
  gravitational wave strain.}
\label{fig:headon_Ham_conv}
\end{figure}

The wave carries away energy from the black hole, so to what degree is
convergence in the waveform reflected in the Hamiltonian (energy)
constraint? Figure~\ref{fig:headon_Ham_conv} plots the Hamiltonian
constraint at the time in which the peak gravitational wave signal
crosses the gravitational wave extraction radius ($t / M \approx 62.9$
and $r_{\text{ext}} / M \approx 44.8$). As with the single puncture
black hole, the Hamiltonian constraint restores its exponential
convergence in the wake of the outgoing gauge pulse. Notice the
Hamiltonian constraint violation in the eighth- and tenth-order finite
difference cases is nearly indistinguishable in the range 
$r / M \gtrsim 7$ due to roundoff error. If
this Hamiltonian constraint violation were to influence the
gravitational waveform convergence, violations at this radius should
impact the gravitational waves at retarded times
$t_{\rm ret} / M \approx 0$ through $52$. However, in that range the 
$|\text{FD 8} - \text{FD 10}|$ and
$|\text{FD 6} - \text{FD 8}|$ curves are easily distinguishable in
Fig.~\ref{fig:gw_FD_convergence}. We conclude that even if the
Hamiltonian constraint is dominated by roundoff error, the
gravitational waveforms may still be convergent.

Based on these results, we conclude that SENR/NRPy+ would be an
excellent tool for studying perturbed black holes.

\section{Conclusion}
\label{sec:Conclusion}

In this work, we extend the reference-metric formulation of the BSSN 
equations pioneered by~\cite{Bonazzola:2004, Brown:2009, Montero:2012, Baumgarte:2013}
to handle Cartesian-, cylindrical-, and spherical-like numerical
grids. At the heart of this strategy, a noncoordinate basis is adopted
to remove from tensorial variables all coordinate singularities that 
arise from the choice of certain bases. 
Treating these singularities analytically, we  
successfully evolve black holes using the moving punctures 
approach~\cite{Alcubierre:2003b, Campanelli:2006, Baker:2006} without
resorting to special integration techniques, and without encountering
numerical instabilities.

We announce a new numerical relativity code package called SENR/NRPy+,
which implements this approach. It is fully open source, open
development, and nonproprietary. NRPy+, written entirely in Python,
converts tensorial expressions and their derivatives in Einstein
notation to optimized C code, representing derivatives with suitable
finite difference approximations. Our current implementation supports
Cartesian-like, spherical-like and cylindrical-like coordinates, but our methods can
be generalized easily for other orthogonal coordinate systems.
SENR contains all of the basic numerical
algorithms needed for a numerical relativity code, making use of the C codes
generated by NRPy+ where complex tensorial expressions are
required.  To the best of our knowledge, this is the first open-source
numerical relativity code that lets the user select from a broad range of curvilinear 
coordinate systems.

The SENR/NRPy+ implementation of the BSSN equations is validated
against two other well established numerical relativity codes. In the
context of a Minkowski spacetime with strong random perturbations, we
achieve roundoff-level agreement with the BMCCM code, which evolves the BSSN equations in
spherical coordinates at fixed fourth-order finite difference
accuracy. We similarly observe excellent agreement between the results
of SENR/NRPy+ and the ETK's McLachlan BSSN
thorn~\cite{Loffler:2012} in the context of a single puncture black
hole evolution. We also show that, for both single and double black
hole spacetimes, the finite difference truncation error
converges with increasing grid resolution at the expected
rate. In addition, we demonstrate exponential convergence of the
error with increasing finite difference order, keeping the grid
resolution constant.

A number of physical diagnostic quantities are implemented in
SENR/NRPy+, including constraint violations and ADM integrals.
For additional tools, we developed an ETK compatibility layer within SENR that
interpolates quantities in the chosen curvilinear coordinate basis to
the Cartesian basis, and onto a Cartesian grid. In this way, SENR can take
direct advantage of the large suite of ETK-based diagnostic utilities,
including, e.g., apparent horizon finders and gravitational wave
diagnostics.  These diagnostics are applied to head-on collisions of
two nonspinning black holes to show that the absolute
difference between gravitational waveforms converges exponentially at
successive finite difference order.

We conclude that our extended formalism for BSSN on arbitrary
coordinate grids as implemented SENR/NRPy+ provides an outstanding
tool for analyzing perturbed black holes without
approximation. Further, the spherical-like coordinate systems adopted
are ideal for gravitational wave extraction and analysis. We next
plan to add coordinate system dynamics and explore bispherical-like
coordinate geometries, so that black hole binaries may be modeled with
extreme efficiency. All simulations displayed in this work can be
performed on aging desktop computers (except for a few of the high
resolution Cartesian runs, which were executed on a desktop with
additional RAM), and given the ability of these coordinate systems to
exploit near-symmetries near compact objects, we anticipate that
SENR/NRPy+ may be the first code to unlock the desktop as a powerful
tool for fully general relativistic gravitational wave astrophysics.

\acknowledgments 
We wish to thank Vassilios Mewes, Manuela Campanelli, and Yosef
Zlochower for many useful discussions. We thank Isabel
Cordero-Carri\'on and Pedro Cerd\'a-Dur\'an for their correspondence
regarding the stability of Runge-Kutta integration. We also thank
Saul Teukolsky for conversations concerning the expected rates of
convergence of our finite difference approximations. Funding for 
computer equipment was provided in part by NSF EPSCoR Grant No.\
OIA-1458952. Some computations were performed on single nodes of West
Virginia University's Spruce Knob high-performance computing
cluster, funded in part by NSF EPSCoR Research Infrastructure
Improvement Cooperative Agreement No.\ 1003907, the state of West
Virginia (WVEPSCoR via the Higher Education Policy Commission), and
West Virginia University. Some computational resources were also 
provided by the Extreme Science and Engineering Discovery 
Environment (XSEDE) Jetstream cloud computing infrastructure at 
Indiana University, via the XSEDE Campus Champion for West Virginia
University account TG-TRA140024. This work was also supported by NSF Grants 
No.\ 1402780 and No.\ 1707526 to Bowdoin College.


\bibliographystyle{apsrev4-1}
\bibliography{references}

\end{document}